\documentclass[]{aastex7}
\usepackage{amsmath}
\usepackage{booktabs,tabularx,siunitx}
\usepackage{caption}
\usepackage{graphicx,subcaption}

\begin{document}

\title{Degradation mechanisms and efficiency of heavily cratered regions on Ceres}

\correspondingauthor{Reem Vitale}

\author[orcid=0000-0002-2497-0349]{Reem Vitale}
\affiliation{Georgia Institute of Technology, Atlanta, Georgia 30332, USA}
\email[]{rghazal3@gatech.edu}  

\author[orcid=0000-0002-1821-5689]{Masatoshi Hirabayashi}
\affiliation{Georgia Institute of Technology, Atlanta, Georgia 30332, USA}
\email[]{thirabayashi@gatech.edu}

\begin{abstract}
Ceres, the dwarf planet in the main asteroid belt, hosts heavily cratered surfaces where craters are continuously eroded mainly due to impact bombardment with a limited influence by non-impact processes. Over continuous bombardment, such regions experience both crater production and erasure, eventually ceasing the crater population growth. This end-state, known as crater equilibrium, provides key information to constrain the mechanisms of crater degradation. The present study applies a recently extended crater equilibrium model to the crater equilibrium features and constrains the conditions for crater degradation on Ceres. We select eight heavily cratered sites as our test locations across four quadrangles (two sites per quadrangle) and collect crater counts using Dawn Framing Camera imagery. All sites exhibit cumulative size-frequency distributions (CSFD) with slopes slightly shallower than a power law of $-2$ at diameters below a few kilometers, strongly suggesting that the tested sites are at crater equilibrium. Our results show that the crater equilibrium state on Ceres resembles that on the Moon but is denser. Performing model fitting with crater counting data under negligible ejecta blanketing for crater erasure, we further show that crater degradation per single crater production on Ceres is comparable to or higher than that on the Moon. Combining this finding and the impact flux on Ceres, which is orders of magnitude higher than that on the Moon, suggests that crater degradation is much more elevated on Ceres than on the Moon, despite its denser crater population.
\end{abstract}



\section{Introduction}

Ceres is a dwarf planet in the main asteroid belt whose surface is heavily cratered due to a long history of impact bombardment \citep{Buczkowski2016, Marchi2016, Strom2018}. Observations from NASA’s Dawn mission have provided high-resolution data revealing that impact craters dominate Ceres's surface morphology, preserving a record of its geological evolution \citep{Russell2016}. A global morphometric analysis showed that craters smaller than 10~km in diameter are predominantly simple or modified simple craters \citep{Buczkowski2016}. Because crater populations encode geologic information such as surface age \citep{Xiao2015}, heavily cratered regions on Ceres offer a valuable case for studying long-term surface processes. The pervasive cratering on Ceres makes it particularly well-suited for investigating the interplay between impact histories and surface modification \citep{Russell2016}.

The high crater density on Ceres suggests prolonged crater production and erasure across much of its surface \citep{Scully2017, Zeilnhofer2020}. The impact flux on Ceres for craters larger than a few hundred m diameter has been orders of magnitude higher than that on the Moon \citep{Hiesinger2016}, because the dwarf planet experienced an extended period of intense impact bombardment \citep{Morbidelli2010, Minton2010}. On the other hand, a numerical study suggests that Ceres has experienced a low micrometeoroid flux, roughly ten times lower than on the Moon and fifty times lower than on Mercury at least at present \citep{Pokorny2021}. In this context, micrometeoroids refer to particles ranging in size from sub-millimeter to millimeter. For the crater diameters analyzed in this study (sub-km or larger in diameter), the relevant impactor population is instead dominated by main-belt asteroids rather than micrometeoroids. At heavily cratered sites on Ceres, impact cratering plays a major role in forming local surface textures with limited cryovolcanic features, in contrast to Ahuna Mons and isolated lobate flows disturbing an otherwise ancient crust \citep{Ruesch2016, Strom2018}. Combining these features infers that heavily cratered regions on Ceres are predominantly affected by continuous impact bombardment of craters larger than a few hundred m diameter, and thus better characterizing crater production and erasure in such regions can constrain impact-driven degradation mechanisms at such locations. 

When crater production continues, exists craters are more likely be erased, eventually ceasing crater production growth. Crater equilibrium represents a steady-state condition where new impacts erase older craters at the same rate they form, yielding a stable size–frequency distribution with a cumulative slope power near $-2$ \citep{Melosh1989}. In this study, we refer to this slope as the crater equilibrium slope. The degree and expression of crater equilibrium may vary among planetary bodies depending on surface composition and degradation processes. For instance, the crater equilibrium state may differ with crater size and surface history \citep{Povilaitis2018}. The crater equilibrium slopes also vary regionally on the Moon, suggesting non-uniform degradation behavior \citep{Xiao2015}. The crater equilibrium slopes for small lunar craters follows a power-law slope near $-2$ \citep{Minton2019}, whereas viscous relaxation and subsurface flow may modify crater equilibrium on icy satellites such as Ganymede and Callisto \citep{Schenk2004}. Crater equilibrium is also scale-variant: small, simple craters and large, complex craters do not reach crater equilibrium under the same conditions \citep{Riedel2020}. On Ceres, while large craters are highly affected by viscous and cryovolcanic relaxation \citep{Bland2016, Fu2017}, smaller craters less than a few km in diameter in heavily cratered regions appear to have reached crater equilibrium.

While empirical characterizations \citep[e.g.,][]{Xiao2015} and numerical modeling \citep[e.g.,][]{Minton2015, Minton2019} are major techniques to investigate crater equilibrium, a semi-analytical approach has applied Poisson's statistics to quantify the same problem \citep{Hirabayashi2017, Hirabayashi2024}. The proposed approach tracks the number of visible craters over crater production and erasure. \cite{Hirabayashi2017} first introduced a degradation parameter, $k$, which represents the efficiency of crater degradation per single crater production. While the details are provided in the following section, the parameter can account for multiple crater degradation mechanisms by considering the relative size of a produced crater to existing craters. In the formulation by \cite{Hirabayashi2017}, the key degradation mechanisms were cookie-cutting, ejecta blanketing, and topographic diffusion. \cite{Hirabayashi2024} later reformulated the degradation parameter and generalized crater equilibrium, defining four crater equilibrium classes. Applying the crater equilibrium model by \cite{Hirabayashi2024} can detail the degradation mechanisms at local sites. 

The purpose of the present study is to (1) demonstrate the application of the established crater equilibrium model developed by \cite{Hirabayashi2024} to empirical data for crater counting for Ceres, and (2) decompose and quantify crater degradation per single crater production. For the first contribution, \cite{Hirabayashi2024} updated the original model proposed by \cite{Hirabayashi2017}. However, \cite{Hirabayashi2024} did not explore the model's applications, and thus it has been unclear how and whether this approach can robustly work when combined with empirical data. For the second contribution, we decompose and quantify crater degradation per crater emplacement. Throughout this work, we obtain crater counting data from eight heavily cratered sites and fit the crater equilibrium model to the empirical data to determine the degradation parameter, which represents the efficiency of crater degradation per crater emplacement. Using the derived quantities, we offer deeper insights into the crater degradation mechanisms at heavily cratered sites on Ceres. To our knowledge, while crater equilibrium has been proposed for heavily cratered regions on Ceres \citep[e.g.,][]{Toyokawa2022}, quantitative characterizations for crater equilibrium and the resulting degradation mechanisms remain limited. The work is organized as follows. Section~\ref{Sec:Methodology} discusses our approach to analyze the crater statistics on Ceres. This section consists of theoretical foundation of our statistical model and crater counting. Section~\ref{Sec:Results} presents the results of applications of the statistical approach. Finally, Section \ref{Sec:Discussion} shares interpretations of our results, focusing on crater equilibrium. 

\section{Geologic activities on Ceres} \label{processes}

In this work, we explore heavily cratered surfaces, where the present statistical framework is applicable. Although non-impact-driven geological processes may influence surface evolution, their effects on crater degradation are likely negligible in the heavily cratered regions considered here. Below, we compile the current understanding of both non-impact and impact-driven processes that could gradually erode surface topography on Ceres.

The first non-impact degradation mechanism is viscous relaxation, which occurs when the ice-rich crust of Ceres slowly flows over time under its weight, causing large craters and basins to structurally relax \citep{Fu2017}. This process likely explains the absence of ancient, very large craters. Simulations predict that at least six or seven basins larger than~400 km should exist on Ceres, yet only two-Kerwan and Yalode-remain visible \citep{Marchi2016}. However, viscous relaxation becomes significant only for craters larger than 20~km in diameter \citep{Otto2019}; craters in the sub-kilometer to 10 km range are largely unaffected.

Another non-impact degradation process includes surface and subsurface flows. Surface flows are morphologically diverse, exhibiting extended material movement along slopes \citep{Buczkowski2016, Schmidt2017, Duarte2019, Chilton2019}. The observed morphological features suggest that they reflect subsurface materials composed of mixed ice and silicates \citep{Duarte2019, Chilton2019}. Some flows may correlate with ejecta generated during crater formation \citep{Hughson2019, Montalvo2022}. On the other hand, subsurface flows refer to the upward migration of brines and cryovolcanic material that modify crater floors and fill depressions. Examples include the bright deposits in Occator \citep{Nathues2017, Nathues2020, Buczkowski2019, Scully2019a, Scully2019b, Quick2019, Schmidt2020} and the dome of Ahuna Mons \citep{Ruesch2016,Sori2017}, both of which indicate resurfacing by cryovolcanic processes in the geologically recent past.  

Impact processes remain the dominant degradation mechanism on Ceres \citep{Marchi2016, Hiesinger2016, Strom2018, Gou2018}. Crater degradation driven by impacts is generally categorized into three mechanisms \citep{Richardson2009, Minton2015, Minton2019, Hirabayashi2017, Hirabayashi2024}, which can be applied to kilometer-sized craters in heavily cratered regions on Ceres. First, cookie-cutting, which occurs when a new crater overlaps the area of any pre-existing craters \citep{Richardson2009}. This process is purely geometric and depends on the relative size of the new craters \citep{Hirabayashi2024}. Second, ejecta blanketing is a process in which debris from an impact site covers older craters and erases them through burial \citep{Minton2019}. However, ejecta blanketing is ineffective when the sizes of source craters are comparable to those of buried craters \citep{Minton2015, Minton2019}. Finally, topographic diffusion, which is the result of numerous small impacts over time that gradually smooth larger craters \citep{Soderblom1970, Fassett2014, Fassett2022, Talkington2022}. The role of micrometeoroid impacts in this process, however, appears to be insignificant at present due to its limited impact flux \citep{Pokorny2021}. Craters are degraded more efficiently by the combined effects of topographic diffusion and, if applicable, ejecta blanketing than through cookie-cutting alone \citep{Minton2019, Hirabayashi2024}.

In addition to the above erosion processes, volatile loss may play a critical role in crater degradation. Water ice exposed in crater walls and rims sublimates over time, weakening the structure and causing wall collapse and infilling. Ice sublimation from crater wall outcrops drives rapid wall retreat and talus formation during the first tens of millions of years after impact cratering, making it the dominant early degradation mechanism \citep{LeBecq2025}. Pole-facing slopes retain ice longer, producing more abundant and steeper talus deposits, whereas equator-facing slopes lose ice more rapidly and stabilize earlier. These deposits evolve through three morphologic stages and disappear in craters older than approximately 3~Ga. Ice-driven wall retreat slightly enlarges craters over time and may contribute to transient water vapor detected near active regions. This mechanism behaves analogously to topographic diffusion \citep{Soderblom1970, Fassett2014, Fassett2022}, modifying diffusive efficiency on Ceres relative to rocky celestial objects like the Moon. 

\section{Methodology}
\label{Sec:Methodology}
\subsection{Crater degradation parameter}
\label{Sec:k}

Crater populations attain a balance between the production of new craters and the degradation of existing ones \citep{Hirabayashi2017}. The crater production process is described by a single-power produced crater cumulative size-frequency distribution (CSFD), $C_t \propto \check D^{-\eta}$, where $\check D$ is the diameter of a produced crater, and $\eta$ is the negative slope power. On the other hand, the rate of crater degradation is governed by the degradation parameter, $k$. This parameter defines a statistical measure of how many pre-existing craters of diameter $D$ are removed by a newly formed crater of diameter $\check D$. The $k$ parameter depends on the relation between $\check D$ and $D$; intuitively, a large new crater can completely obliterate older craters, while a smaller new crater may only partially degrade larger ones \citep{Hirabayashi2017, Hirabayashi2024}. The value of $k$ reflects the dominant physical processes at work. Cookie-cutting yields $k = 1$, particularly when $\check D > D$. Ejecta blanketing on top of cookie-cutting results in $k > 1$, as a large impact with thick ejecta can erase more smaller craters than cookie-cutting alone. In contrast, topographic diffusion leads to $k < 1$ at $\check D \le D$, since numerous small craters are necessary to degrade a large one. 

To capture the described behavior of crater degradation, the degradation parameter is written as follows \citep{Hirabayashi2024}: 
\begin{equation}
    k = 
    \begin{cases}
    \alpha D^{\beta} \left( \dfrac{\check{D}}{D} \right)^{\gamma_1} & \text{if } \check{D} \leq D, \\
    \alpha D^{\beta} \left( \dfrac{\check{D}}{D} \right)^{\gamma_2} & \text{if } \check{D} > D,
    \end{cases}
    \label{Eq:k}
\end{equation}
where $\alpha$ is a scaling factor for the overall degradation efficiency, $\beta$ represents the dependence on the size of the existing crater, and $\gamma_1$ and $\gamma_2$ account for relative-size effects between new and existing craters. $\alpha$, $\gamma_1$, and $\gamma_2$ need to be either zero or positive, while $\beta$ can take any real value. A physical interpretation of these free quantities is that $\alpha$ directly defines the magnitude of the degradation level, and $\beta$ is its size-dependence. $\gamma_1$ characterizes a power of degradation efficiency of larger craters by a single smaller crater emplacement, while $\gamma_2$ represents that of smaller craters by a single larger crater emplacement. Thus, the net effect at $\check D \le D$ corresponds to topographic diffusion per crater emplacement, while that for $\check D > D$ corresponds to cookie-cutting plus ejecta blanketing per crater emplacement. 

\subsection{Crater equilibrium class}

When a cratered surface reaches crater equilibrium, it typically falls into one of four classes, designated as Classes I through IV \citep{Hirabayashi2024}. These classes arise from combinations of the negative crater production CSFD slope power, $\eta$, and the degradation-efficiency parameters, $\gamma_1$ and $\gamma_2$. Class~I represents the most commonly observed crater equilibrium state. This class needs to satisfy $\eta > 2$, i.e., a steep crater production CSFD. The observed crater CSFD in crater equilibrium becomes shallower than the produced crater CSFD, leading to a resulting slope power similar to or slightly lower than $-2$. In addition to the $\eta > 2$ condition, this Class requires to satisfy $0 \le \gamma_2 < \eta - 2 < \gamma_1$ (Section \ref{Sec:equilibrium_state}). All the selected sites in this study are categorized into this class. Class~II arises when the crater production CSFD is steep, and topographic diffusion becomes more efficient than in Class I. Class~III corresponds to the case when a produced crater CSFD shallower than the $-2$-slope power CSFD creates a crater equilibrium CSFD slope with the same slope power. This class has been proposed for the crater population around 100 km diameter on the Moon \citep{Richardson2009, Minton2015}. Finally, Class~IV occurs when the crater production CSFD slope is steep, and both ejecta blanketing and topographic diffusion significantly contribute to crater erasure. This class represents enhanced crater removal relative to Class~I and produces equilibrium slopes that differ from the $-2$-slope power crater equilibrium CSFD.

\subsection{Characterization of crater equilibrium slope}
\label{Sec:equilibrium_state}
This section summarizes the formulation of the Class I crater equilibrium state \citep{Hirabayashi2024}. As described in Section \ref{Sec:k}, the degradation parameter $k$ follows two distinct power-law forms depending on the relative sizes of newly formed and pre-existing craters. The formulation of this parameter follows the updated degradation parameter framework of \citet{Hirabayashi2017}, in which two size conditions are treated separately: (1) when a new crater is smaller than existing craters ($\check{D} \leq D$) and (2) when it is larger than existing craters ($\check{D} > D$). 

The produced crater CSFD, $C_t$, is a function of time $t$ and the produced crater diameter $\check{D}$. This quantity represents the cumulative number of craters larger than $\check{D}$ that have formed per unit area over time. It is expressed as the product of a time-dependent term, $X(t)$, and a size-dependent term, $h(\check{D})$ \citep{Neukum2001}:
\begin{equation}
    C_t(t, \check{D}) = X(t) h(\check{D}).
\end{equation}
When $h(\check{D})$ follows a single power law, $h(\check{D}) = \xi \check{D}^{-\eta}$, we obtain as follows:
\begin{equation}
    C_t = \xi X(t) \check{D}^{-\eta}, \label{Eq:Ct}
\end{equation}
where $\xi$ is a normalization constant, and $X(t)$ is a scaling factor proportional to the integrated cratering rate. For simplicity, we omit the explicit dependence on $(t)$ and write $X(t)$ as $X$. Together, $\xi$ and $X$ describe the crater population state at a given time. When $X(t) = 1$, $\xi$ corresponds to the present-day produced crater CSFD at $\check{D} = 1$~km. The parameters $\xi$ and $\eta$ are determined from empirical fittings. The differential form of the crater production function is then derived by differentiating $C_t$ with respect to $\check{D}$:
\begin{equation}
    \frac{dC_t}{d\check{D}} = -\eta \xi X \check{D}^{-\eta - 1}.
\end{equation}

The observed crater CSFD, $C_c$, evolves according to the balance between the production and removal of craters. It is formulated as an ordinary differential equation, in which the production term uses $dC_t/d\check{D}$, while the removal term integrates the effects of new craters of size $\check{D}$ on existing craters of size $D$. Incorporating the updated degradation parameter $k$, this equation becomes as follows \citep{Hirabayashi2024}:
\begin{equation}
    \frac{dC_c}{dD}=-\frac{\frac{d\dot{C_t}}{dD}}{\frac{\pi}{4}\int^{D_{max}}_{D_{min}}\frac{d\dot{C_t}}{d\check{D}}k\check{D^2}d\check{D}} \left[1-\exp\left\{\frac{\pi}{4}\int^{D_{max}}_{D_{min}}\frac{d{C_t}}{d\check{D}}k\check{D^2}d\check{D}\right\}\right]. \label{Eq:dCc/dD}
\end{equation}
Substituting the expressions for $dC_t / d\check{D}$ and $k$ into the integral and evaluating over the domains $\check{D} \leq D$ and $\check{D} > D$, yields the following form:
\begin{equation}
\frac{\pi}{4} \int_{D_{\min}}^{D_{\max}} \frac{dC_t}{d\check{D}}\, k\, \check{D}^2\, d\check{D}
= \frac{\pi}{4} \eta \xi X \alpha D^{\beta - \eta + 2} \frac{\gamma_1 - \gamma_2}{(-\eta + \gamma_1 + 2)(-\eta + \gamma_2 + 2)}.
\end{equation}
The integration limits are extended to $D_{\min} \rightarrow 0$ and $D_{\max} \rightarrow \infty$. Convergence of this integral depends on the power-law exponents of $\check{D}$. For convergence, the following conditions must be satisfied: $-\eta + \gamma_1 + 2 > 0$ as $\check{D} \to 0$, and $-\eta + \gamma_2 + 2 < 0$ as $\check{D} \to \infty$. These conditions are identical to those of the Class I state, leading to $\gamma_2<\eta-2<\gamma_1$. Because $\gamma_2 \ge 0$, this condition clarifies $0 \le \gamma_2<\eta-2<\gamma_1$

When a cratered surface reaches crater equilibrium, the exponential term in Equation~(\ref{Eq:dCc/dD}) approaches zero, making the equation simple and analytically solvable. The theoretical equilibrium slope is given as follows: 
\begin{equation}
    C_c^\infty = - \frac{4 \Psi}{\alpha \pi (2+\beta)} D^{-2-\beta}, \label{Eq:ctr_eqm}
\end{equation}
where 
\begin{equation} 
    \Psi = \frac{(-\eta + 2 + \gamma_1)(-\eta + 2 + \gamma_2)}{\gamma_1 - \gamma_2} < 0. \label{Eq:Psi_1}
\end{equation}
Equation~(\ref{Eq:ctr_eqm}) provides the crater equilibrium slope by accounting for the crater degradation process described using Equation (\ref{Eq:k}). There are two key features. The first feature is that the slope power, $-2-\beta$, is independent of the produced crater CSFD, indicating that the shallowness or steepness of the crater equilibrium slope is controlled solely by the degradation mechanism. The second feature is that the crater equilibrium CSFD level at $\check D = 1 \text{ km}$, represented by $-4\Psi/\alpha \pi (2 + \beta)$, depends on both $\eta$ and the degradation mechanism. This form, however, is independent of $\xi$, which represents the intensity of the impact flux over time, i.e., surface age. Thus, it demonstrates that our assessment can separate itself from the uncertainties of surface ages to quantify crater degradation. This outcome is reasonable, as the degradation parameter defines how efficiently a single crater emplacement degrades pre-existing craters, which is independent of time. Our approach is to fit this theoretical form to empirical data to determine the degradation parameter $k$ under defined assumptions. 

\subsection{Data collection and processing}
\label{Sec:data_collection}

Manual crater counts are performed using high-resolution images from NASA’s Dawn mission \citep{Russell2011}, obtained through the Small Bodies Image Browser (SBIB) available through the Planetary Science Institute \citep{Nathues2016FC2Ceres}. These images, acquired by the mission’s Framing Camera (FC) developed by the German Aerospace Center (DLR) \citep{Sierks2011}, include orthorectified basemaps from both the High Altitude Mapping Orbit (HAMO) and the Low Altitude Mapping Orbit (LAMO). This dataset provides sufficient resolution to identify craters between 0.4 to 10~km in diameter \citep{Sierks2011}.

Crater mapping is conducted using the Quantum Geographic Information System (QGIS) \citep{QGIS_software}, an open-source GIS platform \citep{Kurt2016}. Dawn FC2 images are imported with the corresponding spatial metadata to ensure accurate area and diameter measurements. The OpenCraterTool plugin \citep{Heyer2023} is used to mark crater locations and record diameters. Each mapped crater is stored as a vector layer with associated attributes and exported as a shapefile for further processing. Craters are identified primarily by rim visibility, with shadows used secondarily to confirm rim curvature and depth. In cases of degraded or partially buried craters, rim continuity and shadow geometry are used to verify crater identity \citep{Robbins2014}. While most craters are circular or near-circular, non-circular craters are fitted using the best circular approximation. Polygonal craters occur on Ceres but are rare in the size range of interest \citep{Buczkowski2016, Bland2016}. Some projected imagery introduces minor shape distortion near image margins, but QGIS minimizes such effects through georeferenced projections, ensuring that measurements are computed in the correct spatial reference frame.

\begin{figure}[ht!]
    \centering
    \includegraphics[width=0.8\linewidth]{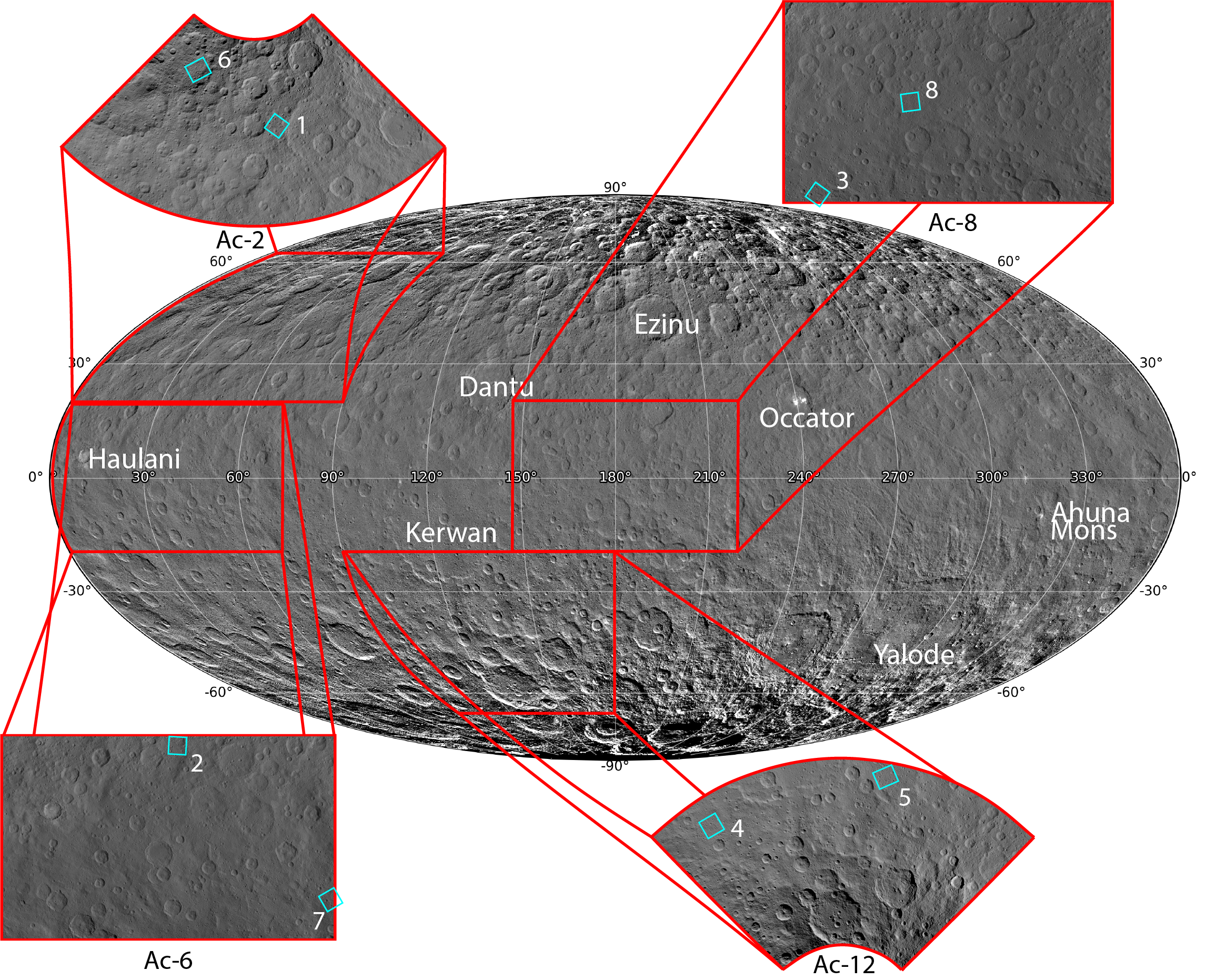}
    \caption{Locations of eight selected crater-counting regions in cyan. Ceres global basemap generated from the Dawn Framing Camera. Data source: NASA/JPL-Caltech/UCLA/MPS/DLR/IDA. The mosaic map is available at the DLR Dawn GIS data portal: \url{https://dawngis.dlr.de/data/Ceres/mosaic_ceres.php}}
    \label{fig:ceres-basemap}
\end{figure}

Eight reference areas are selected, covering four quadrangles for our investigation: Ac-2 \citep{Pasckert2018}, Ac-6 \citep{Krohn2018}, Ac-8 \citep{Frigeri2018}, and Ac-12 \citep{Mest2016}. Figure~\ref{fig:ceres-basemap} shows the locations of these regions on the global Ceres basemap. Investigating multiple quadrangles allows assessment of spatial variations in crater equilibrium. Regions are chosen for high crater density, minimal resurfacing by non-impact processes, and relatively homogeneous terrain. Images are further selected to have solar incidence angles between 42$^\circ$ and 56$^\circ$, balancing illumination and shadowing to enhance crater identification while avoiding extreme lighting biases. Although slightly higher angles ($\sim$74$^\circ$–82$^\circ$) can improve rim detection \citep{Robbins2014}, the moderate angles still provide reliable measurements without the extensive shadow coverage seen at higher values \citep{Robbins2025}. Surface area is measured in QGIS using the \textit{Measure Area} tool, with each image treated as a single counting unit. 

To prepare the data for modeling, the counted craters are then exported in CSV format and binned logarithmically to generate a CSFD for each selected site, following the recommendations of the Crater Analysis Techniques Working Group \citep{CraterAnalysis1979}. Craters smaller than 0.4~km in diameter are excluded to avoid resolution-driven incompleteness, which can artificially flatten the observed CSFD. This cutoff ensures that the fitted slopes reflect genuine equilibrium processes. As a side note, crater populations are sometimes visualized using relative plots (R-plots), which normalize differential crater frequencies by $D^{-3}$ \citep{CraterAnalysis1979, Ivanov2002, Melosh1989}. Although R-plots are useful to visualize deviations, we choose to work directly with CSFDs to preserve the slope information required for fitting the analytical equilibrium model, which is formulated in terms of cumulative rather than differential distributions.

\subsection{Model fitting approach}
\label{Sec:model_fitting}

We apply the analytical model to constrain crater production and degradation. The central idea is to empirically determine the crater equilibrium slope, $C_c^\infty$, and the negative slope power, $\eta$, of the produced crater CSFD, $C_t$, to determine the scalar parameters of the degradation parameter, $k$. Again, Equation (\ref{Eq:ctr_eqm}) depends on $\eta$ but is independent of $\xi$, in $C_t$, indicating that $k$ can be determined without the information of surface age. Before determining $\eta$ for each target site, we identify the observed crater CSFDs from two crater chronology models available for Ceres: the lunar-derived model (LDM) and the asteroid-derived model (ADM) \citep{Hiesinger2016}. Appendix \ref{Sec:crater_production} details these two models. 

For each selected site, we fit the production models with empirical data at larger crater diameters, where the CSFD slopes are steeper and thus can be considered to represent crater production. Given that the LDM and ADM have different slope distributions, and the LDM can offer a better fit than the ADM in most cases, we first fit the LDM with empirical data. A general strategy for the LDM fitting process is to find a crater diameter range that hosts enough crater samples and offers a better fit. We then determine the ADM by assuming the CSFD values of both production models are identical at a diameter of 1 km. We next set a single power law function for $C_t$ by fitting it with a given production model in the considered diameter range, yielding a unique value for $\eta$. We confirm that, in the diameter range fitting, both production functions do not change their slope powers. Therefore, for each production model, the derived $\eta$ values are the same at all the target sites.

Identifying $k$ is achieved through the fitting process applying the crater equilibrium CSFDs. In this process, we determine the scale parameters of $k$, i.e., $\alpha$, $\beta$, $\gamma_1$, and $\gamma_2$ that define the degradation parameter. A key issue, however, is that, given the limited representation of the crater equilibrium features in this study, the present approach cannot determine all the parameters uniquely. To mitigate this issue, we offer two simplifying conditions: $\alpha = 1$, and $\gamma_2 = 0$. The condition $\alpha = 1$ assumes $k = 1$ at $D = \check D$, representing the ideal two-dimensional geometric overlapping, i.e., cookie-cutting. We consider it reasonable. When a produced crater and an existing crater are of the same size, i.e., $\check D = D$, degradation is independent of crater depth, leading to cookie-cutting-like degradation and thus $k=1$ at $\check D = D$. On the other hand, the $\gamma_2 = 0$ condition accounts for a negligible influence of ejecta blanketing on crater degradation. We offer a justification that this condition is reasonable for all the targeted sites in Section \ref{sec:surface_age}. These conditions reduce the number of constants to two: $\beta$ and $\gamma_1$.

The observed crater equilibrium CSFD slope power, $-2-\beta$, can be derived from empirical fitting, uniquely determining $\beta$. The next step is to determine $\gamma_1$. Given that $\gamma_2 = 0$, Equation (\ref{Eq:Psi_1}) becomes as follows:
\begin{equation}
    \Psi = \frac{(-\eta + 2 + \gamma_1)(-\eta + 2)}{\gamma_1}.
\end{equation}
We finally obtain $\gamma_1$ by considering the crater equilibrium CSFD at $D = 1$ km, which is given as follows:
\begin{eqnarray}
    C_c^\infty (D = 1 \: \text{km}) = - \frac{4}{\pi} \frac{(-\eta + 2 + \gamma_1)(-\eta + 2)}{\gamma_1 (2 + \beta)}. \label{Eq:Cc_inf_uni}
\end{eqnarray}

\section{Results}
\label{Sec:Results}

\subsection{Crater counting}
\label{sec:surface_age}

Table \ref{tab:ceres_parameters} shows crater statistics for the eight targeted areas shown in Figure \ref{fig:ceres-basemap}, and Figure \ref{Fig:all-panels} illustrates the observed crater CSFDs for all these sites. Each site is labeled as Reference~$i$ ($i = 1,\ldots,8$). The mapped surface areas range from approximately 1,290 to 2,000~km$^2$, with total crater counts between $\sim$1,100 and $\sim$2,800. The counted craters for all the target sites are available in the Supplemental Materials (Figures S1 through S8). The smallest reliably measured craters have diameters 0.085–0.172~km, depending on image resolution and illumination conditions, while the largest craters range from 4.4 to 15.6~km. This study focuses on crater diameters between 0.4 and 10~km for model fitting and crater equilibrium analysis. We confirm that the geomorphological conditions across all sites are similar, i.e., heavily cratered terrains without evidence of non-impact degradation. The surface ages for each region are also derived by fitting the LDM and ADM to empirical data (Table~\ref{tab:ceres_parameters}). Because the LDM and ADM are not fully consistent, it remains challenging to precisely assess the impact history at each site. This discrepancy has also been noted by \cite{Pasckert2018}, who characterized terrains as ``young" or ``old" rather than relying solely on absolute age estimates. 

While our analysis does not use the surface-age-related parameter, $\xi$, we determine the surface ages for each site using LDM and ADM to confirm consistency with earlier studies. Reference~1 in Ac-2, located just outside Coniraya crater, yields an LDM-based surface age of 1,565~Ma and an ADM-based age of 455~Ma (Figure~\ref{Fig:all-panels}a). The reported LDM age of Coniraya itself is approximately 1,300~Ma, indicating that our derived age is consistent with previous chronology estimates. Across the eight analyzed regions, systematic differences emerge among quadrangles. Regions within Ac-8 (References~3 and~8) and Ac-6 (References~2 and~7) yield older model ages compared to those in Ac-12, consistent with regional age variations previously identified on Ceres \citep{Pasckert2018}. In particular, Reference~3 (Ac-8) exhibits the oldest LDM age of 2,636~Ma (ADM: 768~Ma), while Reference~2 (Ac-6) yields 1,877~Ma (ADM: 545~Ma), placing these terrains among the most ancient in our sample. In contrast, regions within Ac-12 (References~4 and~5) produce younger LDM ages of 962~Ma and 1,361~Ma (ADM: 279~Ma and 395~Ma, respectively), again consistent with the presence of smooth plains and resurfacing deposits reported in the northwestern sector of this quadrangle \citep{Mest2016}.

While counting craters across all the target sites, we also examine whether clear ejecta blanketing is present. All the target sites are heavily bombarded, though the observed craters do not exhibit obvious ejecta- or ray-like morphologies surrounding them. Ejecta blanketing from much larger craters can influence much smaller ones. If this process affects a local site, the observed crater CSFD may exhibit a complex profile that deviates from a single-power law \citep{Minton2019}. We examine the crater equilibrium CSFDs for all the targeted sites and confirm that they follow single-power laws (Figure \ref{Fig:all-panels}), excluding the critical influence of  ejecta blanketing on crater degradation. In this analysis, we do not account for ejecta blanketing from large craters outside each target area. If ejecta blanketing from these external craters is significant, a target site may be resurfaced completely, resetting the local impact history. We also carefully inspect whether such ejecta blanketing partially affects our target sites. However, we do not observe any recognizable geologic contrasts across the target sites, indicating no complication of ejecta-blanketing-driven crater degradation.

\subsection{Model fitting with empirical data}
\label{Sec:fitting}

\begin{figure}
    \centering
    \includegraphics[width=\linewidth]{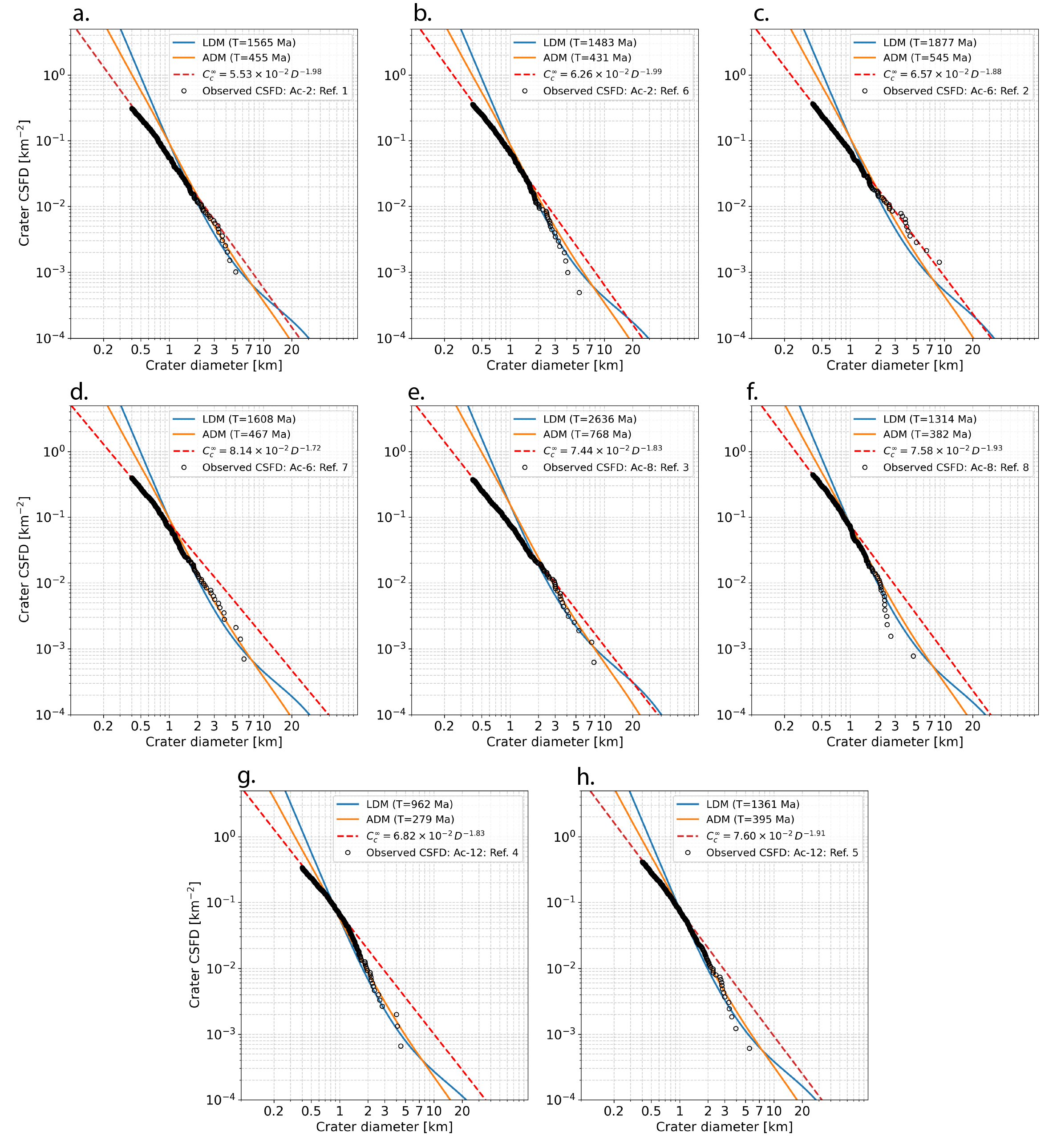}
    \caption{Observed and fitted crater size-frequency distributions (CSFDs) for eight counting areas on Ceres. Each panel corresponds to each reference: a.~Reference 1 (Ac-2); b.~Reference 6 (Ac-2); c.~Reference 2 (Ac-6);  d.~Reference 7 (Ac-6); e.~Reference 3 (Ac-8); and f.~Reference 8 (Ac-8); g.~Reference 4 (Ac-12); and h.~Reference 5 (Ac-12). Black circles denote the empirical data from our crater counting, blue and orange curves show the modeled lunar- and asteroid-derived chronology fits (LDM and ADM, respectively), and red dashed lines are the equilibrium state, $C_c^\infty$.}
    \label{Fig:all-panels}
\end{figure}

We follow the approach described in Section \ref{Sec:model_fitting} to determine the constants ($\beta$ and $\gamma_1$) in the degradation parameter, $k$, under $\alpha = 1$ and $\gamma_2 = 0$. The shallow slopes are interpreted as crater equilibrium, and their slope powers appear to be similar across all the target sites (Figure \ref{Fig:all-panels}). The crater equilibrium slope powers range between $-1.72$ and $-1.99$. Again, the crater equilibrium slopes at all the sites can generally be described using single power laws. The observed crater equilibrium slope power, $-2-\beta$, is applied to determine $\beta$. $\beta$ ranges between $-0.28$ and $-0.01$ for all eight heavily cratered sites (Table \ref{tab:ceres_parameters}). All the derived $\beta$ values are negative and show no clear correlation with  surface age. 

We determine the $\gamma_1$ and $\eta$ parameters for all target sites. Applying the fitting process discussed earlier, we determine the negative produced crater CSFD slope powers as $\eta = 3.2$ for the LDM and $\eta = 2.4$ for the ADM. We then substitute these values of $\eta$ into Equation~(\ref{Eq:Cc_inf_uni}), computing $\gamma_1$ for each site. The resulting $\gamma_1$ values are listed in Table~\ref{tab:ceres_parameters}, ranging from 1.29 to 1.33 for the LDM and from 0.51 to 0.57 for the ADM. All the derived \(\gamma_1\) values are positive, indicating that crater degradation efficiency increases with crater size. Comparison between the two production models further shows that a steeper production slope corresponds to a higher $\gamma_1$, giving a less influence of topographic diffusion on crater degradation. We also confirm that all the selected sites are in the Class I crater equilibrium state. The produced crater CSFDs for all the target sites satisfy $\eta > 2$. Furthermore, the Class I condition requires $\eta - 2 < \gamma_1$. For the LDM cases, $\eta - 2 = 1.2$, while $\gamma_1$ is between 1.29 and 1.33. This condition meets the $\eta - 2 < \gamma_1$ condition. For the ADM cases, $\eta - 2 = 0.4$, and $\gamma_1$ ranges from 0.51 to 0.57, again satisfying the Class I condition.

We also compute the break crater diameter, $D_b$, i.e., the diameter at which the observed CSFD begins to systematically deviate from the steeper production slope and instead follows the shallower equilibrium slope. The break diameter is evaluated in the diameter-CSFD space on a log scale. For each selected site, we confirm that $D_b$ lies along both $C_t$ and $C_c^\infty$. The inferred break diameters, which range between approximately 1 and 2~km across the analyzed regions, are consistent with the diameter range reported by an earlier study that identified deviations from the production function and interpreted them as equilibrium on Ceres \citep{Toyokawa2022}.

Finally, the present work applies the crater equilibrium model developed by \cite{Hirabayashi2024}, who updated the mathematical form of the $k$ parameter. \cite{Hirabayashi2017} initially proposed the concept of quantifying crater degradation by introducing the first version of the $k$ parameter. However, \cite{Hirabayashi2024} pointed out that the form introduced by \cite{Hirabayashi2017} would encounter singularity problems if the parameters for the $k$ parameter are not selected properly. The present study applies the improved form to eight heavily cratered sites on Ceres and confirms that the expression of the $k$ parameter does not encounter any singularity issues. This finding indicates the robustness of the applied crater equilibrium model for various crater equilibrium conditions.

\section{Discussion}
\label{Sec:Discussion}

\subsection{Crater degradation efficiency}
\label{Sec:degradation}

This study studied  eight heavily cratered regions on Ceres to examine crater equilibrium and degradation. Crater equilibrium at all the selected sites is categorized as Class I. The production profile is steeper than a $-2$-power crater production CSFD. This type of crater equilibrium is considered to be the most common on planetary surfaces, at least on the Moon \citep{Minton2019}. The degradation parameter, $k$, is the efficiency of crater degradation by single crater production and thus independent of the number of produced craters. In the Class I state, $\beta$ and $\gamma_1$ are not functions of $\xi$, which represents surface age if $X = 1$ (Equation~(\ref{Eq:Cc_inf_uni})). From the present results, we confirm the time-independent nature of these quantities. In other words, $\beta$ and $\gamma_1$ are similar, independent of location and surface age on Ceres. Considering that cookie-cutting and topographic diffusion play a role in crater degradation, this section presents our interpretation of crater degradation efficiency based on the two constants, $\beta$ and $\gamma_1$, of the derived $k$ parameter.

In the following discussion, we compare our results with the crater equilibrium slopes derived by \cite{Hartmann1984} and \cite{Minton2019}. \cite{Hartmann1984} reported a crater equilibrium slope for diameters between 0.06 and 32~km on the Moon, as well as on Phobos and Deimos, as $C_c^\infty = 0.046 D^{-1.83}$. According to the lunar crater production model \citep{Neukum2001}, the crater diameter range considered by \cite{Hartmann1984} spans steeper and shallower production slopes, with a transition at 1~km. Thus, their crater equilibrium state likely results in a combination of Class I for diameters less than 1~km and Class III for those larger than this value. In the following discussion, therefore, we only refer to lunar craters up to 1~km when citing \cite{Hartmann1984}. On the other hand, at the Apollo~15 landing site in Mare Imbrium on the Moon, the equilibrium slope for craters less than 0.2~km in diameter was reported as $C_c^\infty = 0.02 D^{-1.8}$ \citep{Minton2019}.

The $\beta$ parameter represents the size-dependent effect on crater degradation in $k$. This quantity can be determined from the crater equilibrium slope power, $-2-\beta$. At the target sites, the crater equilibrium slope power ranges from $-1.72$ to $-1.99$, varying by only about 10\%. This range is consistent with that observed on the Moon \citep{Hartmann1984, Minton2019}. Our derived $\beta$ values, thus, range between $-0.28$ and $-0.01$ (Table \ref{tab:ceres_parameters}). No clear correlation was observed between $\beta$ and surface age, suggesting that these quantities are independent. The negative $\beta$ values at all the selected sites indicate that $k$ becomes lower (higher) than in the size-independent case ($\beta = 0$) when a target crater's size is larger (smaller). This means that larger craters are more resistant to degradation and thus take longer to be erased than predicted under the ideal size-independent condition. This interpretation is consistent with earlier crater counting work for topographic diffusion of large craters on the Moon \citep{Talkington2022}. 

The elevated effect of crater degradation per crater emplacement is further discussed using the $\gamma_1$ parameter. This parameter represents how effectively a single smaller crater emplacement can degrade larger craters, i.e., topographic diffusion per crater emplacement. A higher $\gamma_1$ means lower topographic diffusion per crater emplacement. For Ceres, it ranges from 0.51 to 0.57 for the ADM and from 1.29 to 1.33 for the LDM. On the other hand, the $\gamma_1$ value is 1.27 for lunar craters for the \cite{Hartmann1984} case, while it is 1.23 for those up to 1 km in Mare Imbrium \citep{Minton2019}. The lunar $\gamma_1$ values are obtained by applying the lunar production model \citep{Neukum2001}, which is provided in Appendix \ref{Sec:crater_production} and Table~\ref{tab:aj_coefficients}. Similar to the $\gamma_1$ values for Ceres, we fit the $C_t$ function with the lunar production model at a crater diameter of 1 km. Around this diameter, the $C_t$ function can be described using a single-power law \citep{Neukum2001}, leading to a constant value of $\eta = 3.2$. Considering a wider crater size range for this fit does not change the $\eta$ value, making our outcomes robust. The LDM-based $\gamma_1$ value (1.29-1.33) is comparable to the lunar $\gamma_1$ value (1.23-1.27), whereas the ADM-based $\gamma_1$ value (0.51-0.57) is lower than the lunar value. This finding suggests that topographic diffusion per crater emplacement on Ceres may be similar to (or $\sim$10\% lower in the LDM than) that on the Moon, or $\sim$5 times higher in the ADM, if pre-existing craters are 10 times larger than a newly produced crater. If pre-existing craters are 100 times larger than the produced crater, topographic diffusion per crater emplacement on Ceres may be just $\sim$30\% lower in the LDM or $\sim$26 times higher in the ADM than on the Moon. These conditions suggest that topographic diffusion (per single crater emplacement) on Ceres is comparable or elevated compared to that on the Moon. 

We further provide insight into the net level of crater degradation, accounting for both topographic diffusion and cookie-cutting. The net effect of each degradation mechanism can be the product of the contribution of a single crater production event to degradation multiplied by the total number of produced craters. Cookie-cutting is purely geometrical and depends only the size of a newly produced crater. Meanwhile, topographic diffusion per crater emplacement on Ceres is comparable or higher than that on the Moon. On the other hand, the impact flux on Ceres is orders of magnitude higher than on the Moon. Therefore, combining these factors suggests that the net level of crater degradation on Ceres is orders of magnitude higher than on the Moon.

\subsection{Crater equilibrium density}

This section introduces the concept of crater saturation to describe how densely craters are distributed at crater equilibrium on Ceres. We introduce two saturation concepts: geometric saturation and cookie-cutter saturation \citep{Melosh1989, Hirabayashi2024}. 

Geometric saturation represents a hypothetical, nonphysical upper limit, where craters are ideally placed at the highest packing conditions, allowing overlap in their placement and size distribution \citep{Melosh1989, Minton2019, Hirabayashi2024}:
\begin{eqnarray}
    {C_{gs}^{\infty} = 1.54 D^{-2}}
\end{eqnarray}
where $C_{gs}^{\infty}$ is the geometric saturation slope, with units of km$^{-2}$. Neglecting variations in the crater equilibrium slope power, the $C_c^\infty(D = \text{1 km})$ level relative to the geometric saturation condition, hereafter referred to simply as the saturation level, suggests that the target sites reach crater equilibrium at less than $5$\% of geometric saturation. This is within the reported saturation range on the Moon, which is between 1 and 10\% \citep{Xiao2015}. However, the geometric saturation concept becomes unreliable when the derived crater equilibrium slope power deviates from the ideal $-2$ value. It also relies on assumptions that are not physically realistic \citep{Hirabayashi2024}.

The cookie-cutter saturation, in contrast, results from the process of cookie-cutting, especially when large craters overlap existing ones \citep{Hirabayashi2024}. To model this condition, we extend the formulation of \cite{Hirabayashi2024}, leaving $\beta$ as a free parameter while imposing the following constraints: $\gamma_1 \rightarrow \infty$, $\gamma_2 = 0$, and $\alpha = 1$. These correspond to zero topographic diffusion ($\gamma_1 \rightarrow \infty$), zero ejecta blanketing ($\gamma_2 \rightarrow 0$), and the absence of three-dimensional effects on cookie-cutting ($\alpha = 1$). Under these assumptions, the cookie-cutter saturation slope is given as follows:
\begin{eqnarray}
    C_{cs}^\infty = \frac{4 (\eta - 2)}{\pi (2 + \beta)} D^{ -2 - \beta}
    \label{Eq:C_s^inf_new}
\end{eqnarray}
This expression depends on $\eta$; when $\eta = 3$ and $\beta = 0$, it reduces to the form presented by \cite{Hirabayashi2024}. Using the derived $\beta$ and $\eta$ values for each site yields a cookie-cutter saturation level, expressed as $C_c^\infty (D = \text{1 km}) / C_{cs}^\infty (D = \text{1 km})$, ranging between 7-10\% for the LDM, and 21-30\% for the ADM across all the references. For comparison, applying Equation (\ref{Eq:C_s^inf_new}) yields a cookie-cutter saturation level of 2.3-5.5$\%$ on the Moon \citep{Hartmann1984, Minton2019}. Hence, crater equilibrium  on Ceres is denser than that on the Moon by more than a factor of a few. 

\subsection{Geological conditions of heavily cratered sites on Ceres}

In this study, we selected heavily cratered terrains to quantify the crater equilibrium state made primarily by impact cratering processes. Applying a statistical crater equilibrium model, the previous sections offered insights into a statistical sense of crater equilibrium on Ceres: higher topographic diffusion per crater emplacement and denser crater equilibrium. This section offers our geological interpretations of the crater equilibrium state at heavily cratered sites on Ceres.

One interpretation of the observed dense crater equilibrium state on Ceres is that the widespread presence of highly cohesive and porous materials helps preserve sharp crater morphologies despite elevated topographic diffusion. Ceres's surface is relatively uniformly covered by phyllosilicates (hydrated silicates) \citep{Ammannito2016}, derived from materials formed at depth \citep{Castillo-Rogez2019}. Topographic studies pointed out that Ceres may exhibit globally higher angles of repose than the Moon or Mars \citep{Ermakov2019}, possibly reflecting regional-scale geology \citep{Ermakov2019}. Experimental work suggested that Ceres's surface materials are highly porous and cohesive, as mixtures containing water ice are exposed to intense sublimation processes \citep{Schroder2021}. Highly cohesive materials tend to resist granular deformation and therefore remain less dynamic than cohesionless ones. Thus, the high angles of repose reported by \citet{Ermakov2019} are consistent with this experimental findings of \citet{Schroder2021}. We interpret that the dense crater population observed at crater equilibrium, despite high crater degradation, reflects the ability of cohesive, porous surface materials to preserve key crater morphologies, enabling degraded craters to remain recognizable in imagery.  

The selected heavily cratered sites lack the existence of water in surface/subsurface layers. The presence of water in Ceres's surface layers has been reported in multiple remote-sensing studies \citep{Combe2016, Platz2016}. However, exposed surface ices are thermally unstable under solar illumination \citep{Landis2017, Formisano2018}. Subsurface water ice \citep{Bland2016, Fu2017} may also enhance crater degradation, as localized melting can mobilize materials and produce fluidized flows \citep{Schmidt2017, Duarte2019, Chilton2019}. Impact-driven ejecta may also be fluidized by a similar mechanism, such as ice melting \citep{Hughson2019, Montalvo2022}. However, our target sites did not indicate such water-driven morphological features. The absence of these signatures, combined with the presence of highly cohesive and porous materials, suggests that heavily cratered regions likely formed after the loss of surface water. While the timing and processes that led to this surface condition remain unconstrained, they may relate to the formation of ammoniate phyllosilicates, the ubiquitous mineralogical component on Ceres \citep{Ammannito2016}, and its timescale, connected with the formation of this object \citep{Singh2021}. 

Secondary cratering may contribute to Ceres's crater population. Secondaries form from ejecta produced by primary impacts and may resemble primaries in morphology \citep{McEwen2006, Bierhaus2018}. Because their morphologies can be similar, distinguishing them is often difficult unless they form obvious clusters \citep{Gou2018, Schenk2020}. Earlier studies have generally found that the produced secondary crater CSFD is steeper than that of primaries, implying a higher $\eta$ value for secondaries. Using $\eta = 4$ as an example \citep{Bierhaus2018}, the cookie-cutter saturation level is estimated at 4-6\%, the lowest among the $\eta$ values considered in this study, and comparable to that on the Moon. Similarly, the corresponding $\gamma_1$ values range from 2.0 to 2.1, the highest among our samples. Nevertheless, because secondary cratering also plays an important role on the Moon, secondary production alone cannot properly explain the differences in degradation efficiency between Ceres and the Moon. 

Finally, we briefly mention about the relationship between our results and the findings of \citet{Costello2021}, who suggested that Ceres experiences less impact gardening than the Moon. At first glance, our finding that the net level of crater degradation is higher on Ceres than on the Moon appears inconsistent with their conclusion. \cite{Costello2021} reported the level of impact gardening at 1 Ga. Over this period, the impact flux on Ceres is estimated to be roughly 70 times higher under the LDM and 240 times higher under the ADM than that on the Moon (Table \ref{tab:aj_coefficients}). If impact flux alone drives gardening, its intensity should scale proportionally with impact flux \citep{Hirabayashi2018, Costello2020}. The apparent discrepancy may arise from their crater counting approach limited to diameters larger than 1 km \citep{Gou2018, Michael2020}, possibly underestimating the impact flux. Our crater counting data suggest that the observed crater CSFD at 1 km in diameter can be in crater equilibrium, as shown in Figure \ref{Fig:all-panels}. The break diameters at some sites exceed 1 km (Table \ref{tab:ceres_parameters}). The LDM and ADM have production slopes of $-3.2$ and $-2.4$, respectively, both steeper than the observed crater equilibrium slopes with slope powers of $-1.83$ to $-1.98$. However, we agree that low-speed impacts on Ceres may make impact gardening inefficient \citep{Costello2021}. Combining our findings of elevated crater degradation with those of \cite{Costello2021} on less impact gardening at deeper depths suggests intensive shallow degradation mechanisms that quickly erase existing craters. 

\section{Conclusion}
This study investigated crater degradation efficiency on Ceres by applying a statistical model that can analyze crater equilibrium by introducing a scalar quantity called the degradation parameter. This parameter comprises multiple constants representing three degradation effects: cookie-cutting, topographic diffusion, and ejecta blanketing. In the degradation parameter, $\beta$ quantifies the deviation of the crater equilibrium CSFD slope power from $-2$, while $\gamma_1$ measures how efficiently a single crater emplacement degrades larger craters, thus topographic diffusion per crater emplacement. Our statistical analysis at eight heavily cratered sites on Ceres revealed a $\beta$ range between $-0.01$ and $-0.28$, which is consistent with that on the Moon. The negative $\beta$ value suggests inefficient degradation of larger craters compared to smaller ones. Applying the lunar-based crater production model yielded a $\gamma_1$ range between 1.29 and 1.33, while using the lunar-based model gave a $\gamma_1$ range between 1.23 and 1.27, and using the asteroid-belt-based model yields a $\gamma_1$ range from 0.51 to 0.55. The derived $\gamma_1$ values are thus comparable to or lower than those on the Moon, suggesting that topographic diffusion per crater emplacement on Ceres is comparable or elevated compared to that on the Moon. With these findings, a much higher impact flux on Ceres than on the Moon leads to a higher net crater degradation level on Ceres than on the Moon. We determined a cookie-cutter saturation level of 7-30\% on Ceres, depending on a crater production model, in contrast to 2.3-5.5\% on the Moon. Our interpretation is that Ceres's cohesive and porous materials preserve sharp crater morphologies, allowing degraded craters to remain identifiable in imagery despite elevated degradation activity.

\section{Acknowledgments}
This work is performed under support of 80NSSC24K0802. 

\appendix

\section{Crater production and chronology model for Ceres}
\label{Sec:crater_production}
The statistical model requires a calibrated crater production function to determine how craters accumulate over time. While there are many impact flux models in the main asteroid belt \citep{Bottke2005, Minton2010, Nesvorny2017}, an earlier study proposed two models for Ceres: the Lunar-Derived Model (LDM) and the Asteroid-Derived Model (ADM) \citep{Hiesinger2016}. 

The LDM scales the lunar production function to Ceres by adjusting fluxes using crater counts from well-dated lunar reference surfaces and accounting for differences in heliocentric distance \citep{Schmedemann2014}. This scaling accounts for the lower intrinsic collision probability and reduced average impact velocities in the main belt compared to near-Earth space. This approach assumes that the relative shape of the production function remains the same as on the Moon, with the overall flux level modified for Ceres's orbital environment. However, because impacts on the Moon and Earth are dominated by near-Earth objects, while impacts on Ceres come primarily from main-belt asteroids, the LDM may misrepresent the true projectile population.

In contrast, the ADM derives the crater production function directly from the observed size–frequency distribution (SFD) of main-belt asteroids \citep{OBrien2014, Hiesinger2016}. In this approach, the measured asteroid SFD is first converted into an impactor population for Ceres by accounting for orbital position in the main belt and the intrinsic collision probability between asteroids and Ceres. This distribution is then transformed into a crater SFD using impact scaling laws that link impactor size to crater size. This method is physically grounded in the current asteroid population and dynamical environment, without assuming that the shape of the production function matches that of the Moon.

Both models, therefore, have shortcomings. The LDM may import an inappropriate size distribution from the lunar–terrestrial environment, while the ADM depends on poorly constrained small-body statistics and dynamical models~\citep{Marchi2016}. Neither fully captures the unique conditions of Ceres in the main asteroid belt. This issue causes geologic studies (particularly crater statistics) to report geologic ages with intentional ambiguity \citep{Pasckert2018}, though some studies report geologic ages from both models \citep{Hiesinger2016, Scully2018}. Thus, earlier reports tend to describe surface ages as ``young" or ``old" without stating their geologic ages. 

Below, we discuss our implementation of crater production to the developed model. Both LDM and ADM are incorporated into the Neukum chronology model \citep{Neukum2001}. It describes the cumulative number of craters larger than diameter or equal to $D$ per km$^2$, denoted as $N (\ge D)$:
\begin{equation}
    \log_{10} N (\ge D) = \sum_{j=0}^{11} a_j (\log_{10} D)^j,
\end{equation}
where $a_j$ are the coefficients of a 11th order polynomial, found in Table~\ref{tab:aj_coefficients}. The cumulative number of craters larger than 1~km formed over time $T$ (in Ga) on the Moon is expressed as follows:
\begin{equation}
    M(\ge 1 \text{ km}, T) = c_1  \left[\exp(c_2  T) - 1\right] + c_3 T.
\end{equation}
The crater production flux for $\ge D$ is obtained by scaling this quantity using the relative abundance $N(\ge D) / N(\ge 1 \text{ km})$, yielding the following form:
\begin{equation}
    F(\ge D, T) = M(\ge 1 \text{ km}, T) \cdot \frac{N(\ge D)}{N(\ge 1 \text{ km})}.
\end{equation}
Table~\ref{tab:aj_coefficients} lists the chronology coefficients, $c_1$, $c_2$, and $c_3$, originally derived for the Moon \citep{Neukum2001} and those for both LDM and ADM on Ceres \citep{Hiesinger2016}. 

\begin{table}[ht!]
\centering
\caption{Polynomial coefficients, $a_i (i = 0, ..., 11)$, and chronology coefficients, $c_i (i = 1, 2, 3)$, for the lunar \citep{Neukum2001} and Ceres \citep{Hiesinger2016} production functions. The coefficients of the lunar-derived model (LDM) come from \cite{Hiesinger2016}, and those of the asteroid-derived model (ADM) are derived from a fitting process using the Neukum model \citep{Neukum2001}. Both LDM and ADM are only applicable to the impact history up to 4.1 Ga \citep{Hiesinger2016}.}
\begin{tabular}{lccc}
\hline
\textbf{Coefficient} & \textbf{Moon} & \textbf{Ceres (LDM)} & \textbf{Ceres (ADM)} \\
\hline
\hline
Production model \\
\hline
$a_0$  & $-2.5489$       & $-2.8502$ & $-1.0640$ \\
$a_1$  & $-2.9794$       & $-3.2567$ & $-2.7017$ \\
$a_2$  & $0.42605$       & $0.48621$ & $-0.10316$ \\
$a_3$  & $0.32288$       & $0.60437$ & $0.43539$ \\
$a_4$  & $-0.030823$     & $0.07478$ & $0.47841$ \\
$a_5$  & $-0.022295$     & $-0.18704$ & $-0.64544$ \\
$a_6$  & $0.019473$      & $-0.072493$ & $-0.073077$ \\
$a_7$  & $-0.022278$     & $0.02096$ & $0.44950$ \\
$a_8$  & $-0.0085611$    & $0.013396$ & $-0.36665$ \\
$a_9$  & $0.0053854$     & $-0.00072764$ & $0.16317$ \\
$a_{10}$ & $0.00087331$  & $-0.00080732$ & $-0.038597$ \\
$a_{11}$ & $-0.0003887$  & $1.5624 \times 10^{-5}$ & $0.0036806$ \\
\\
Chronology model \\
\hline 
$c_1$ & $5.44 \times 10^{-14}$ & $3.79 \times 10^{-12}$ & $2.09 \times 10^{-6}$ \\
$c_2$ & $6.93$ & $6.93$ & $2.57$ \\
$c_3$ & $8.38 \times 10^{-4}$ & $5.84 \times 10^{-2}$ & $0.201$ \\
\hline
\end{tabular}
\label{tab:aj_coefficients}
\end{table}

\newpage
\clearpage
\pdfpagewidth=11in
\pdfpageheight=8.5in

\begin{minipage}{\textheight}
\centering
\captionof{table}{Fitted parameters and image metadata for selected regions on Ceres. $C_c^\infty (D = 1 \text{ km})$ is the crater equilibrium CSFD at $D = 1$ km, $\beta$ represents deviation from the -2 crater equilibrium slope power, $\eta$ is the produced crater CSFD, $\gamma_1$ is the scale parameter representing topographic diffusion for the degradation parameter, $k$, and $C_{cs}^\infty$ is the cookie-cutter saturation CSFD at $D = 1$ km.}
\label{tab:ceres_parameters}

\begin{tabular}{lcc cc cc cc}
\toprule

\textbf{Quadrangle}
& \multicolumn{2}{c}{\textbf{Ac-2}}
& \multicolumn{2}{c}{\textbf{Ac-6}}
& \multicolumn{2}{c}{\textbf{Ac-8}}
& \multicolumn{2}{c}{\textbf{Ac-12}} \\

\cmidrule(lr){2-3}
\cmidrule(lr){4-5}
\cmidrule(lr){6-7}
\cmidrule(lr){8-9}

\textbf{Reference}
& 1 & 6 & 2 & 7 & 3 & 8 & 4 & 5 \\

\midrule

\textbf{Image ID} & 16063122646 & 16119125155 & 16127014244 & 16112065337 & 16167094521 & 16127072636 & 16058131352 & 16109133002 \\
\textbf{Area [km\textsuperscript{2}]} & 1,976 & 2,021 & 1,412 & 1,430 & 1,597 & 1,290 & 1,508 & 1,634 \\
\textbf{Counted craters} & 2,806 & 1,521 & 1,800 & 1,767 & 1,109 & 1,804 & 1,095 & 1,570 \\
\textbf{Central latitude [$^\circ$]} & 42.32 & 53.60 & 18.83 & -13.19 & -22.22 & 0.50 & -25.66 & -24.12 \\
\textbf{Central longitude [$^\circ$]} & 53.76 & 17.72 & 38.97 & 72.52 & 148.08 & 171.02 & 101.64 & 146.91 \\
\textbf{Minimum crater diameter [km]} & 0.085 & 0.149 & 0.086 & 0.113 & 0.172 & 0.118 & 0.149 & 0.121 \\
\textbf{Maximum crater diameter [km]} & 15.55 & 5.407 & 8.841 & 6.215 & 7.752 & 4.677 & 4.444 & 5.488 \\
\textbf{Solar incidence angle [$^\circ$]} & 55.83 & 58.14 & 41.7  & 42.34 & 56.51 & 39.83 & 52.46 & 46.4 \\
\textbf{LDM-based surface age [Ma]} & 1,565 & 1,483 & 1,877 & 1,608 & 2,636 & 1,314 & 962 & 1,361 \\
\textbf{ADM-based surface age [Ma]} & 455 & 431 & 545 & 467 & 768 & 382 & 279 & 395 \\
\textbf{Break diameter $D_b $ [km]} & 1.3 & 1.9 & 1.4 & 0.9 & 2.0 & 0.8 & 1.2 & 1.3 \\

\\
\multicolumn{9}{l}{\textbf{Crater equilibrium state}} \\
\midrule

\textbf{$C_c^\infty(D=\text{1 km})$ [km$^{2+\beta}$]} & 0.055 & 0.066 & 0.065 & 0.067 & 0.074 & 0.071 & 0.065 & 0.080 \\
\textbf{$\beta$ [-]} & $-0.02$ & $-0.01$ & $-0.12$  & $-0.28$ & $-0.17$  & $-0.07$ & $-0.17$  & $-0.09$ \\

\\
\multicolumn{9}{l}{\textbf{LDM-based fitting}} \\
\midrule

\textbf{$\eta$ [-]} & 3.2 & 3.2 & 3.2 & 3.2 & 3.2 & 3.2 & 3.2 & 3.2 \\
\textbf{$\gamma_1$ [-]} & 1.29 & 1.31 & 1.31 & 1.30 & 1.32 & 1.32 & 1.30 & 1.33 \\
\textbf{$C_{cs}^\infty(D=\text{1 km})$ [km$^{2+\beta}$]} & 0.77 & 0.77 & 0.81 & 0.89 & 0.83 & 0.79 & 0.83 & 0.80 \\
\textbf{$C_c^\infty(D=\text{1 km})/C_{cs}^\infty(D=\text{1 km})$ [-]} & 0.071 & 0.085 & 0.081 & 0.075 & 0.089 & 0.090 & 0.078 & 0.10 \\

\\
\multicolumn{9}{l}{\textbf{ADM-based fitting}} \\
\midrule

\textbf{$\eta$ [-]} & 2.4 & 2.4 & 2.4 & 2.4 & 2.4 & 2.4 & 2.4 & 2.4 \\
\textbf{$\gamma_1$ [-]} & 0.51 & 0.54 & 0.53 & 0.52 & 0.54 & 0.55 & 0.52 & 0.57 \\
\textbf{$C_{cs}^\infty(D=\text{1 km})$ [km$^{2+\beta}$]} & 0.26 & 0.26 & 0.27 & 0.30 & 0.28 & 0.26 & 0.28 & 0.27 \\
\textbf{$C_c^\infty(D=\text{1 km})/C_{cs}^\infty(D=\text{1 km})$ [-]} & 0.21 & 0.25 & 0.24 & 0.23 & 0.27 & 0.27 & 0.23 & 0.30 \\

\bottomrule
\end{tabular}
\end{minipage}
\clearpage

\pdfpagewidth=8.5in
\pdfpageheight=11in

\newpage

\section{Supplementary Material\label{section:supplementary_material}}

\begin{figure}[h!]
    \centering
    \includegraphics[width=\linewidth]{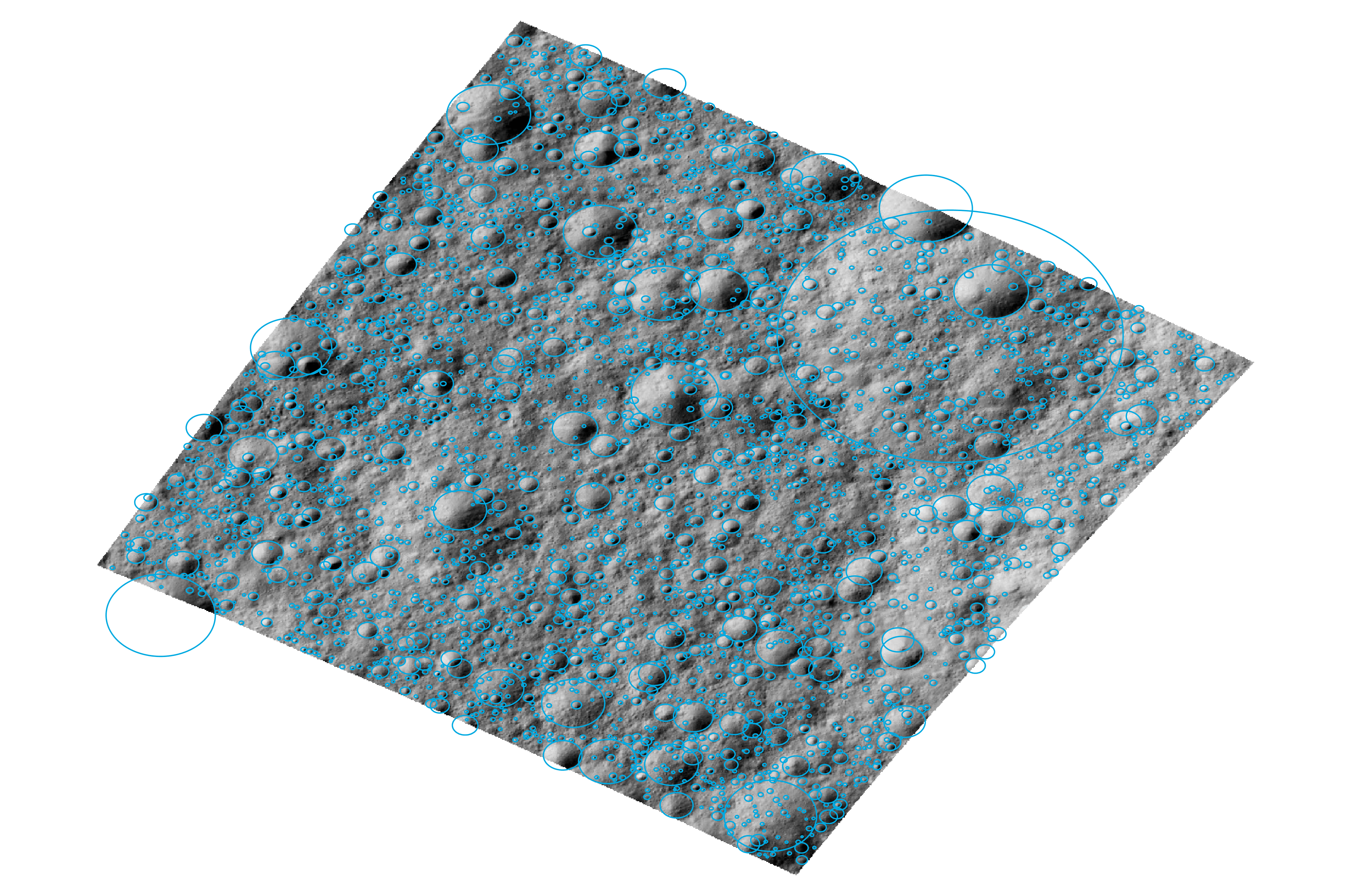}
    \caption{Mapped crater population for Reference~1 (Ac-2). 
    Dawn Framing Camera image ID 16063122646 centered at 42.32$^\circ$ latitude and 53.76$^\circ$ longitude, covering an area of 1,976~km$^2$. 
    A total of 2,806 craters were identified and measured. 
    The grayscale mosaic used for crater identification is shown with all craters included in the statistical analysis outlined in blue. 
    Crater diameters were measured rim-to-rim and used to construct the cumulative crater size--frequency distribution (CSFD).}
    \label{fig: Ac-2-ref1}
\end{figure}

\begin{figure}
    \centering
    \includegraphics[width=\linewidth]{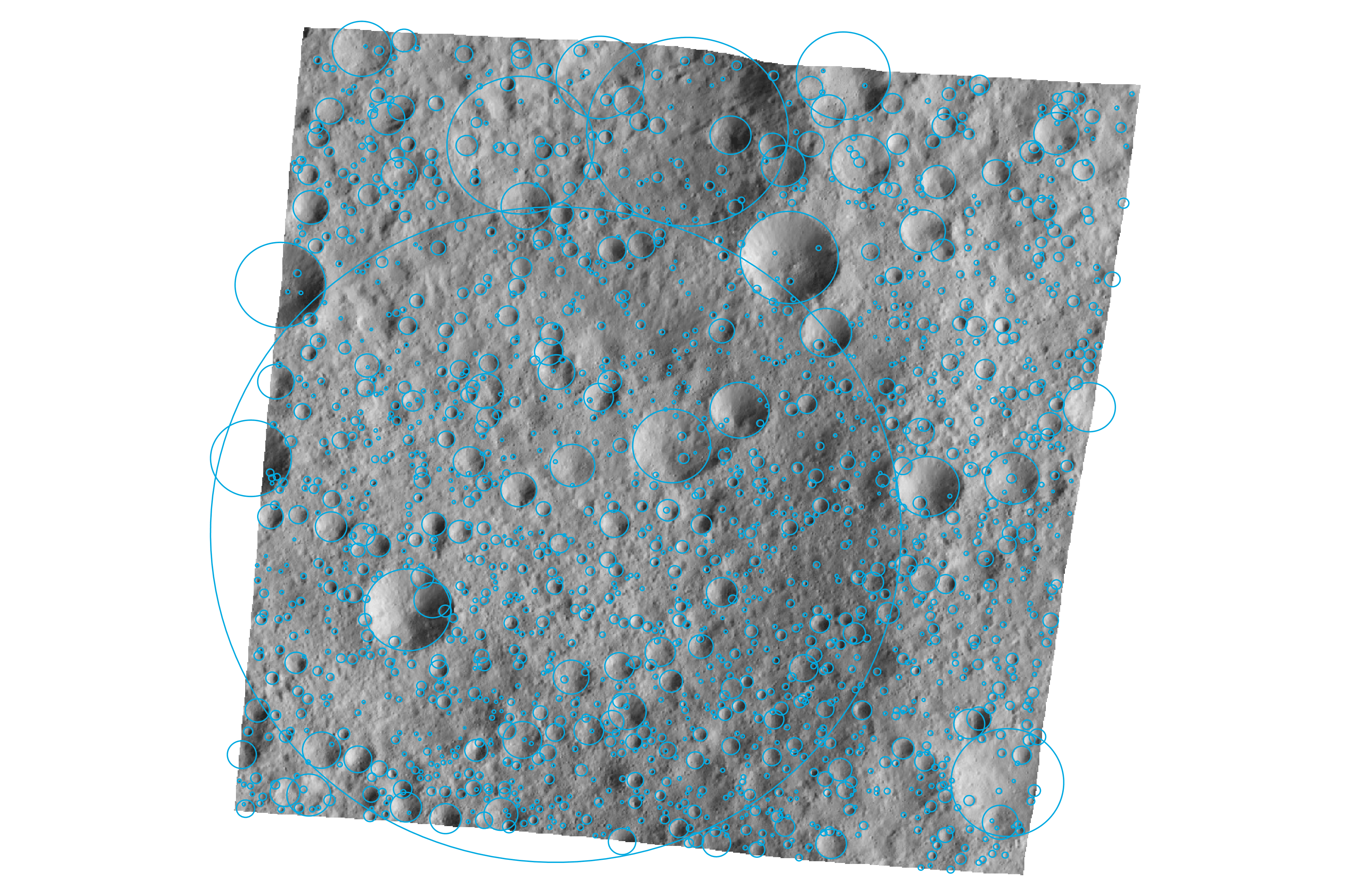}
    \caption{Mapped crater population for Reference~2 (Ac-6). 
    Dawn Framing Camera image ID 16127014244 centered at 18.83$^\circ$ latitude and 38.97$^\circ$ longitude, covering an area of 1,412~km$^2$. 
    A total of 1,800 craters were identified and measured. 
    The grayscale mosaic used for crater identification is shown with all craters included in the statistical analysis outlined in blue. 
    Crater diameters were measured rim-to-rim and used to construct the cumulative crater size--frequency distribution (CSFD).} 
    \label{fig: Ac-6-ref2}
\end{figure}

\begin{figure}
    \centering
    \includegraphics[width=\linewidth]{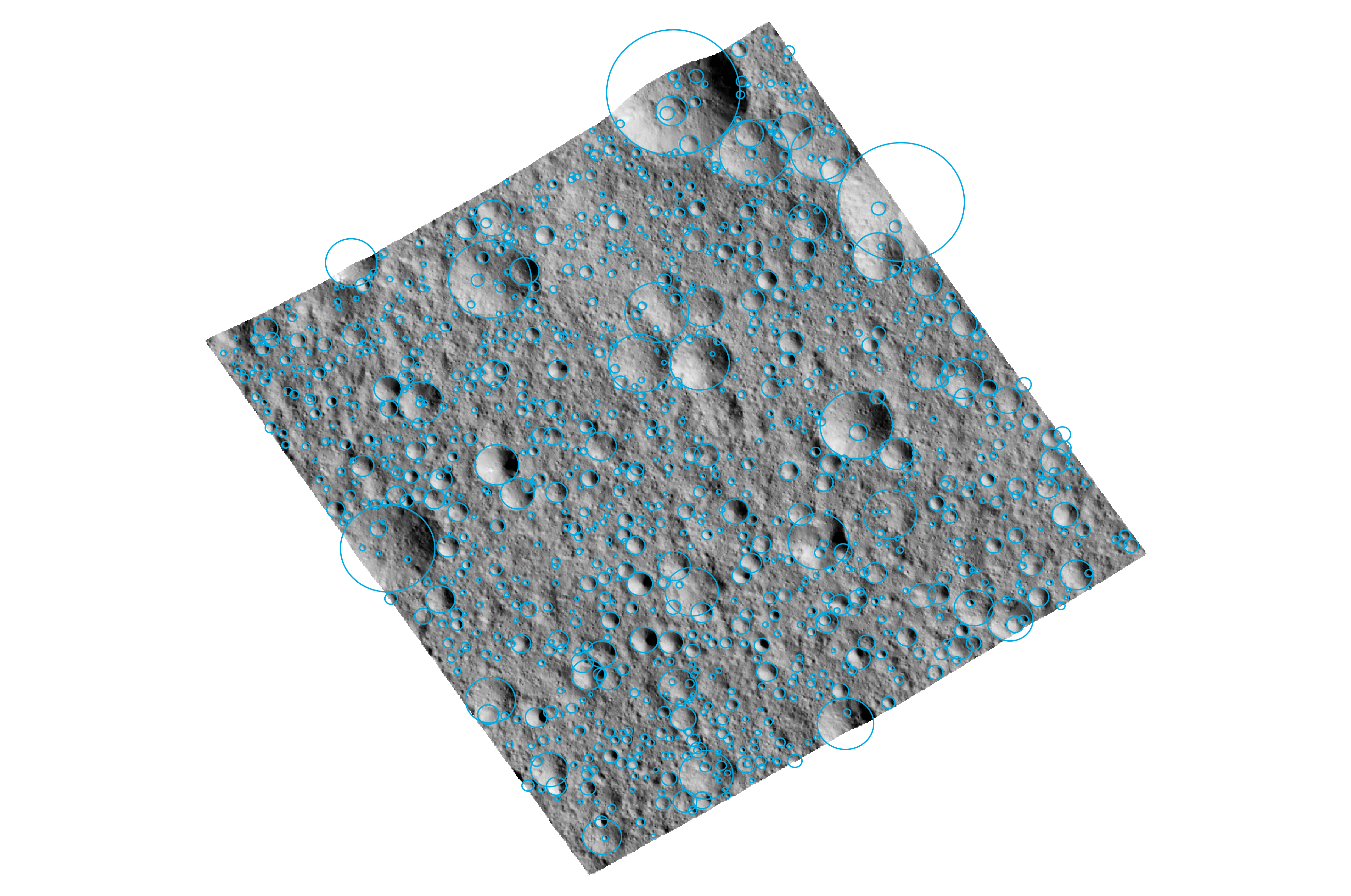}
    \caption{Mapped crater population for Reference~3 (Ac-8). 
    Dawn Framing Camera image ID 16167094521 centered at $-22.22^\circ$ latitude and 148.08$^\circ$ longitude, covering an area of 1,597~km$^2$. 
    A total of 1,109 craters were identified and measured. 
    The grayscale mosaic used for crater identification is shown with all craters included in the statistical analysis outlined in blue. 
    Crater diameters were measured rim-to-rim and used to construct the cumulative crater size--frequency distribution (CSFD).}
    \label{fig: Ac-8-ref3}
\end{figure}

\begin{figure}
    \centering
    \includegraphics[width=\linewidth]{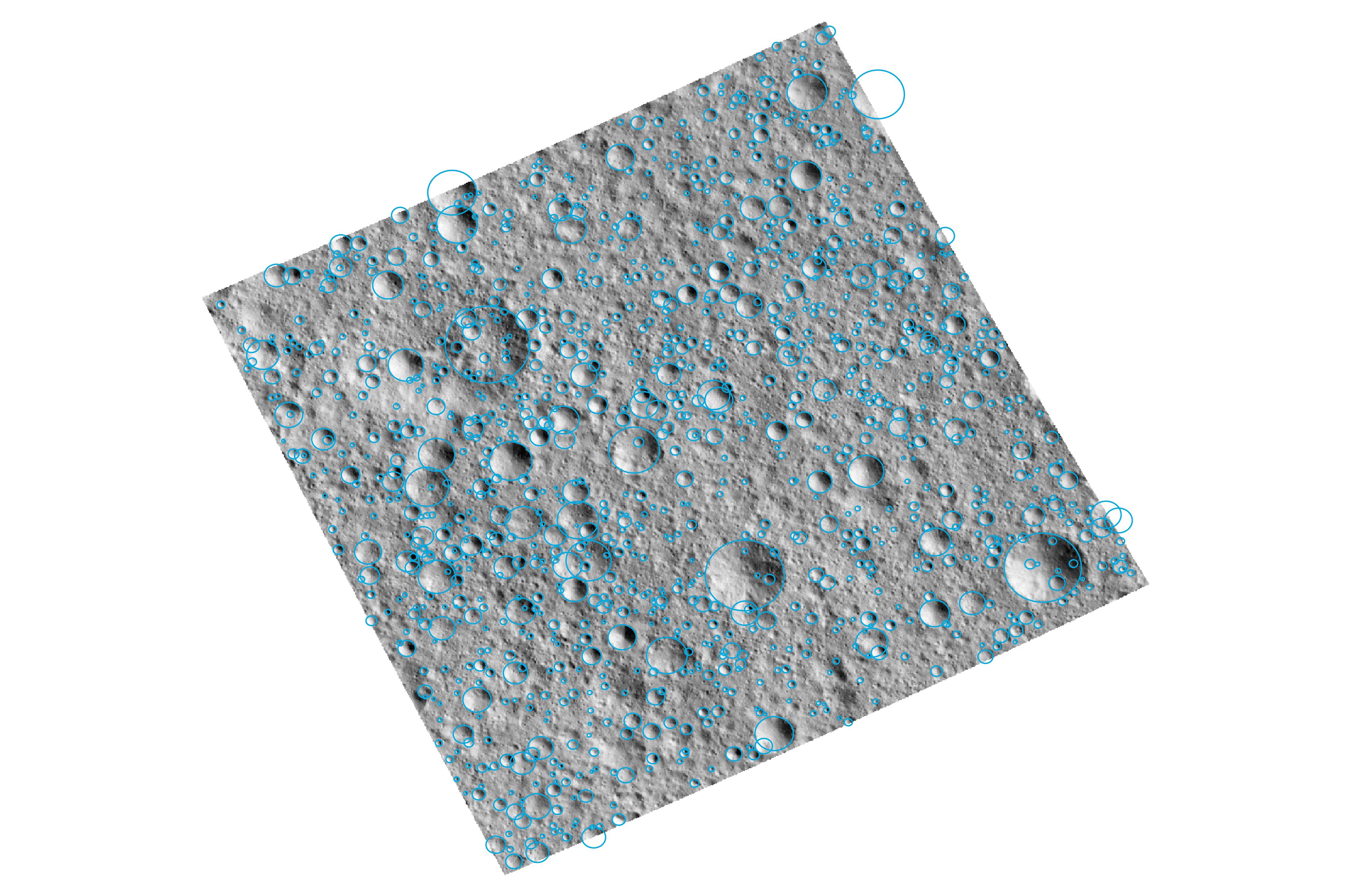}
    \caption{Mapped crater population for Reference~4 (Ac-12). 
    Dawn Framing Camera image ID 16058131352 centered at $-25.66^\circ$ latitude and 101.64$^\circ$ longitude, covering an area of 1,508~km$^2$. 
    A total of 1,095 craters were identified and measured. 
    The grayscale mosaic used for crater identification is shown with all craters included in the statistical analysis outlined in blue. 
    Crater diameters were measured rim-to-rim and used to construct the cumulative crater size--frequency distribution (CSFD).}   
    \label{fig: Ac-12-ref4}
\end{figure}

\begin{figure}
    \centering
    \includegraphics[width=\linewidth]{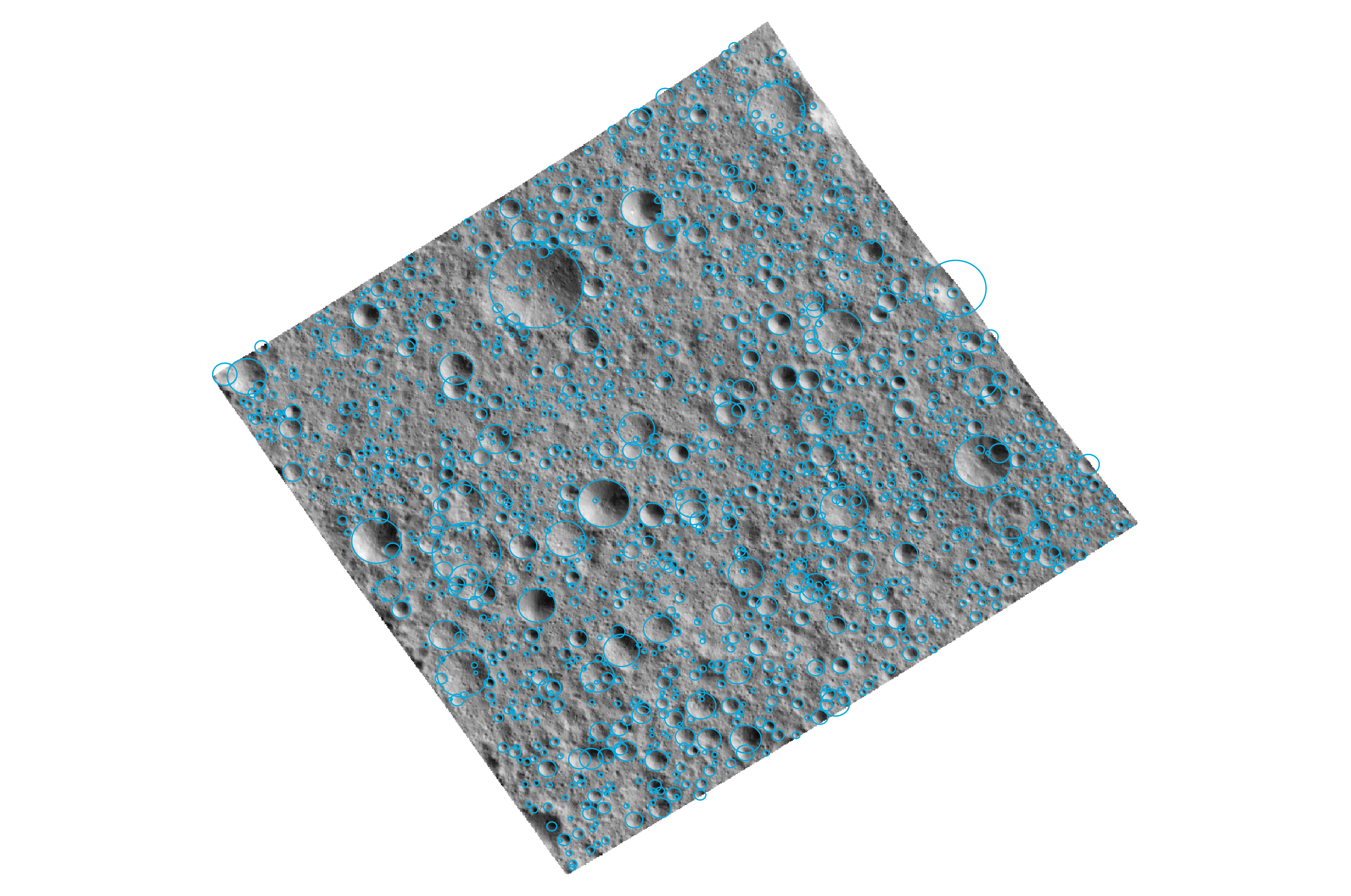}
    \caption{Mapped crater population for Reference~5 (Ac-12). 
    Dawn Framing Camera image ID 16109133002 centered at $-24.12^\circ$ latitude and 146.91$^\circ$ longitude, covering an area of 1,634~km$^2$. 
    A total of 1,570 craters were identified and measured. 
    The grayscale mosaic used for crater identification is shown with all craters included in the statistical analysis outlined in blue. 
    Crater diameters were measured rim-to-rim and used to construct the cumulative crater size--frequency distribution (CSFD).} 
    \label{fig: Ac-12-ref5}
\end{figure}

\begin{figure}
    \centering
    \includegraphics[width=0.8\linewidth]{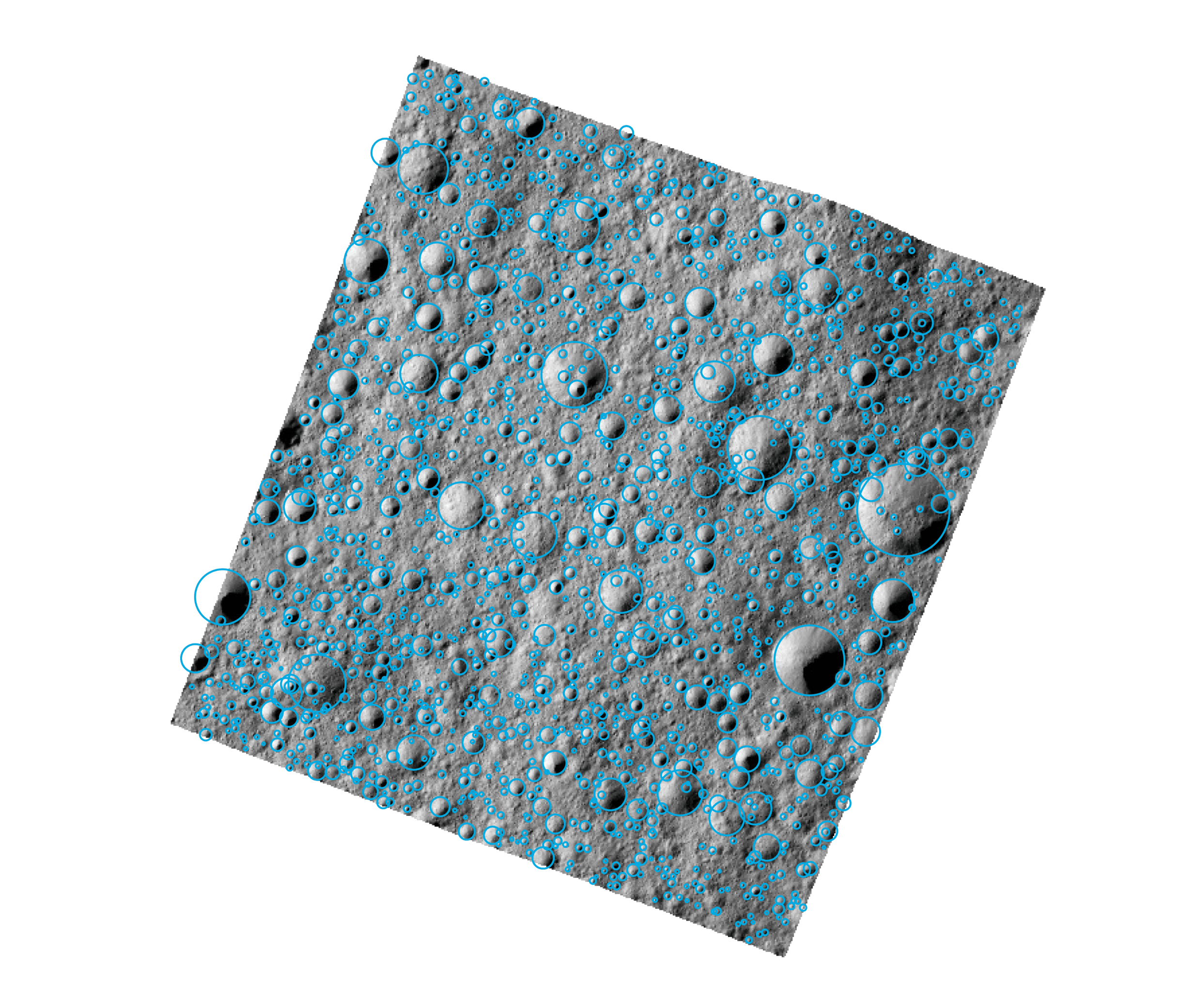}
    \caption{Mapped crater population for Reference~6 (Ac-2). 
    Dawn Framing Camera image ID 16119125155 centered at 53.60$^\circ$ latitude and 17.72$^\circ$ longitude, covering an area of 2,021~km$^2$. 
    A total of 1,521 craters were identified and measured. 
    The grayscale mosaic used for crater identification is shown with all craters included in the statistical analysis outlined in blue. 
    Crater diameters were measured rim-to-rim and used to construct the cumulative crater size--frequency distribution (CSFD).}    
    \label{fig: Ac-2-ref6}
\end{figure}

\begin{figure}
    \centering
    \includegraphics[width=\linewidth]{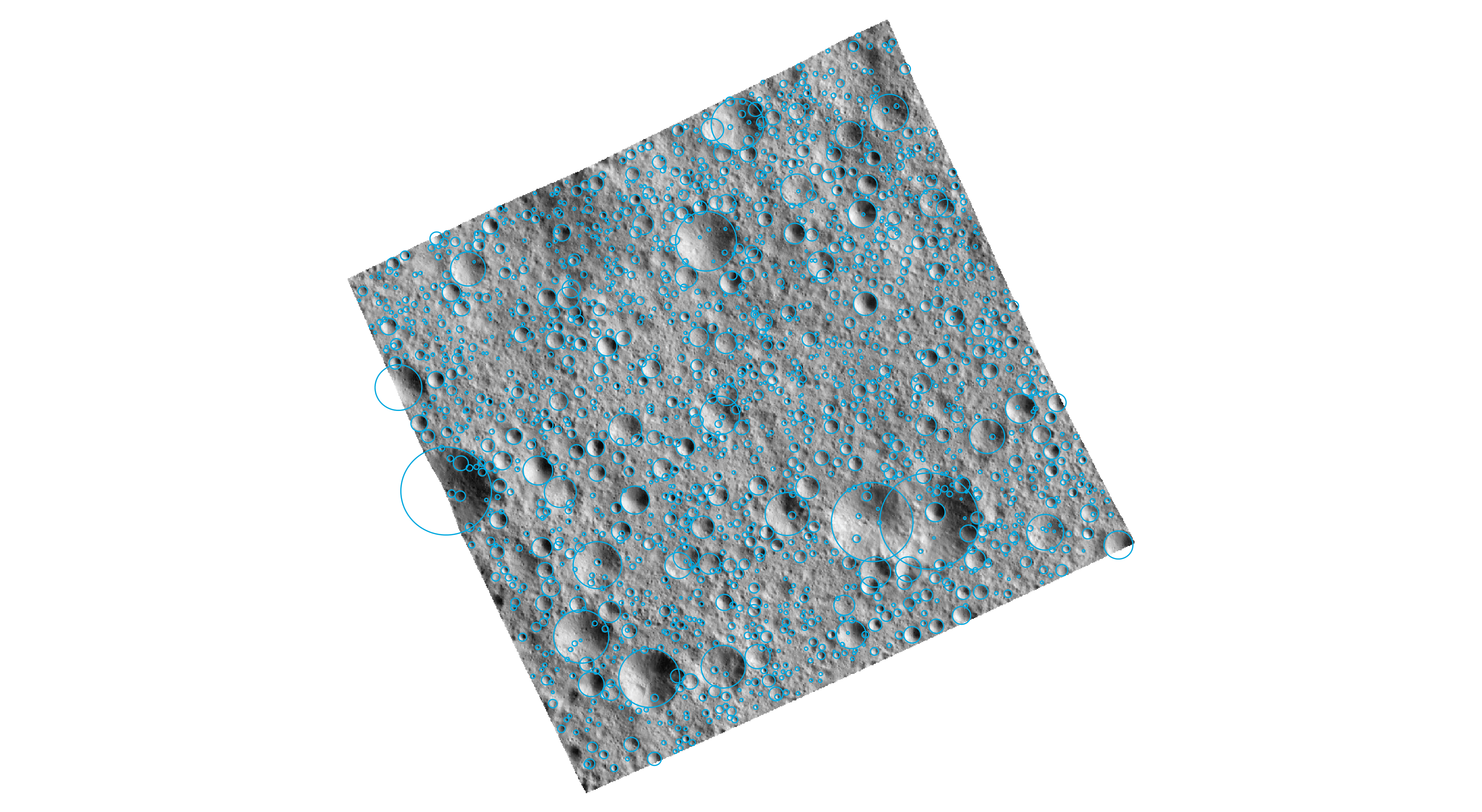}
    \caption{Mapped crater population for Reference~7 (Ac-6). 
    Dawn Framing Camera image ID 16112065337 centered at $-13.19^\circ$ latitude and 72.52$^\circ$ longitude, covering an area of 1,430~km$^2$. 
    A total of 1,767 craters were identified and measured. 
    The grayscale mosaic used for crater identification is shown with all craters included in the statistical analysis outlined in blue. 
    Crater diameters were measured rim-to-rim and used to construct the cumulative crater size--frequency distribution (CSFD).}  
    \label{fig: Ac-6-ref7}
\end{figure}

\begin{figure}
    \centering
    \includegraphics[width=\linewidth]{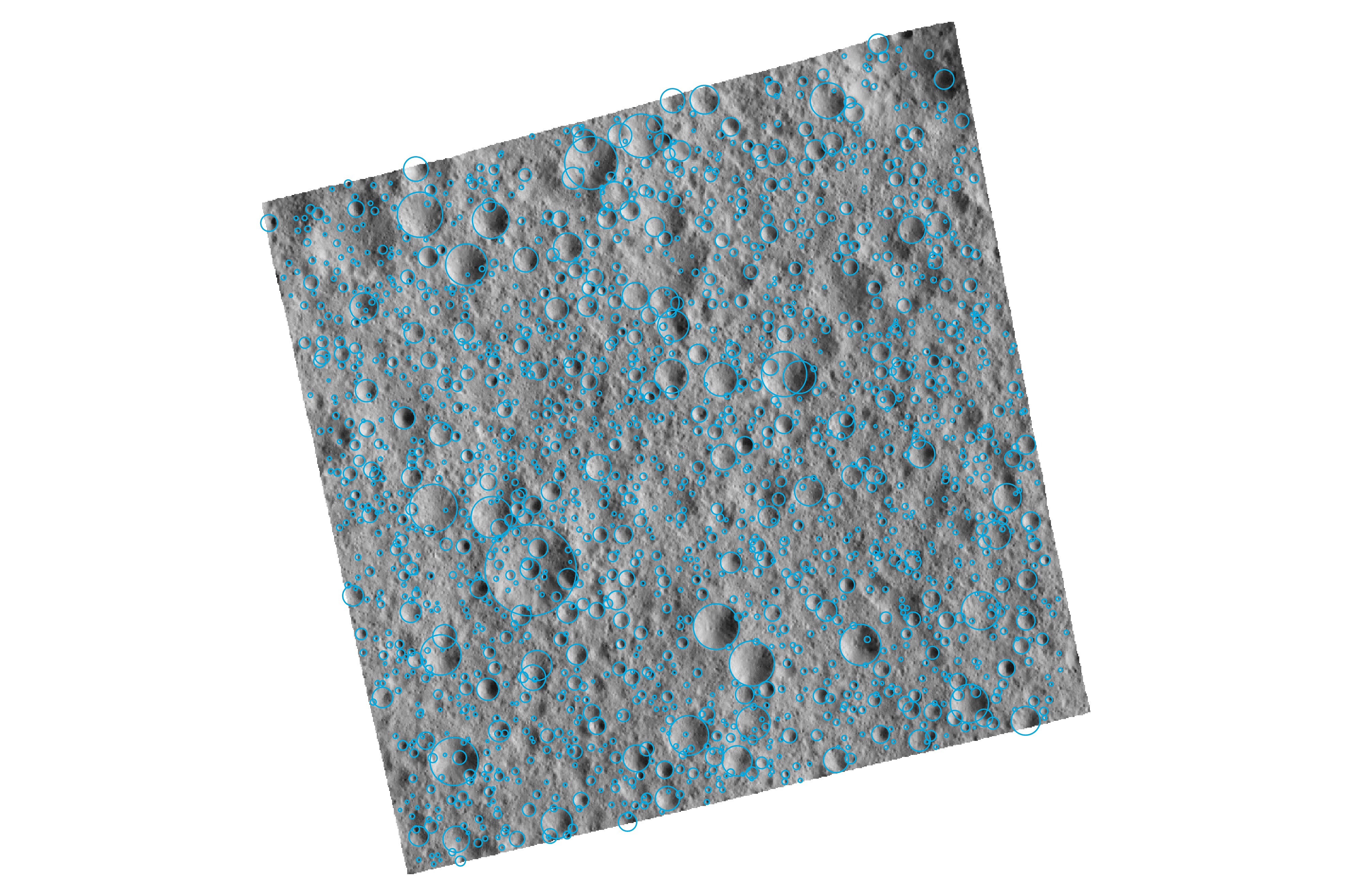}
    \caption{Mapped crater population for Reference~8 (Ac-8). 
    Dawn Framing Camera image ID 16127072636 centered at 0.50$^\circ$ latitude and 171.02$^\circ$ longitude, covering an area of 1,290~km$^2$. 
    A total of 1,804 craters were identified and measured. 
    The grayscale mosaic used for crater identification is shown with all craters included in the statistical analysis outlined in blue. 
    Crater diameters were measured rim-to-rim and used to construct the cumulative crater size--frequency distribution (CSFD).}
    \label{fig: Ac-8-ref8}
\end{figure}

This section includes supplementary crater-counting figures (Figures \ref{fig: Ac-2-ref1}--\ref{fig: Ac-8-ref8}). Other supplementary materials for this manuscript are available at \dataset[DOI: 10.5281/zenodo.18040632]{https://doi.org/10.5281/zenodo.18040632}, which include a global map of Ceres with the counting regions, analysis code, and parameter tables.

\clearpage


\begin{thebibliography}{}
\expandafter\ifx\csname natexlab\endcsname\relax\def\natexlab#1{#1}\fi
\providecommand{\url}[1]{\href{#1}{#1}}
\providecommand{\dodoi}[1]{doi:~\href{http://doi.org/#1}{\nolinkurl{#1}}}
\providecommand{\doeprint}[1]{\href{http://ascl.net/#1}{\nolinkurl{http://ascl.net/#1}}}
\providecommand{\doarXiv}[1]{\href{https://arxiv.org/abs/#1}{\nolinkurl{https://arxiv.org/abs/#1}}}

\bibitem[{E. Ammannito {et~al.}(2016)Ammannito, DeSanctis, Ciarniello, Frigeri,
  Carrozzo, Combe, Ehlmann, Marchi, McSween, Raponi, Toplis, Tosi,
  Castillo-Rogez, Capaccioni, Capria, Fonte, Giardino, Jaumann, Longobardo,
  Joy, Magni, McCord, McFadden, Palomba, Pieters, Polanskey, Rayman, Raymond,
  Schenk, Zambon, \& Russell}]{Ammannito2016}
Ammannito, E., DeSanctis, M.~C., Ciarniello, M., {et~al.} 2016,
  \bibinfo{title}{Distribution of phyllosilicates on the surface of Ceres,}
  Science, 353, aaf4279, \dodoi{10.1126/science.aaf4279}

\bibitem[{N.~L. Becq {et~al.}(2025)Becq, Conway, Jabaud, Tobie, \&
  Artoni}]{LeBecq2025}
Becq, N.~L., Conway, S., Jabaud, B., Tobie, G., \& Artoni, R. 2025,
  \bibinfo{title}{A new model of crater degradation on Ceres involving ice
  sublimation and talus formation,} Icarus, 428, 116353,
  \dodoi{10.1016/j.icarus.2024.116353}

\bibitem[{E.~B. Bierhaus {et~al.}(2018)Bierhaus, McEwen, Robbins, Singer,
  Dones, Kirchoff, \& Williams}]{Bierhaus2018}
Bierhaus, E.~B., McEwen, A.~S., Robbins, S.~J., {et~al.} 2018,
  \bibinfo{title}{Secondary craters and ejecta across the solar system:
  Populations and effects on impact-crater–based chronologies,} Meteoritics
  \& Planetary Science, 53, 638, \dodoi{10.1111/maps.13057}

\bibitem[{M.~T. Bland {et~al.}(2016)Bland, Raymond, Schenk, Fu, Kneissl,
  Pasckert, Hiesinger, Preusker, Park, Marchi, King, Castillo-Rogez, \&
  Russell}]{Bland2016}
Bland, M.~T., Raymond, C.~A., Schenk, P.~M., {et~al.} 2016,
  \bibinfo{title}{Composition and structure of the shallow subsurface of Ceres
  revealed by crater morphology,} Nature Geoscience, 9, 538,
  \dodoi{10.1038/ngeo2743}

\bibitem[{W.~F. Bottke {et~al.}(2005)Bottke, Durda, Nesvorný, Jedicke,
  Morbidelli, Vokrouhlický, \& Levison}]{Bottke2005}
Bottke, W.~F., Durda, D.~D., Nesvorný, D., {et~al.} 2005,
  \bibinfo{title}{{Linking the collisional history of the main asteroid belt to
  its dynamical excitation and depletion},} Icarus, 179, 63,
  \dodoi{10.1016/j.icarus.2005.05.017}

\bibitem[{D.~L. Buczkowski {et~al.}(2016)Buczkowski, Schmidt, Williams, Mest,
  Scully, Ermakov, Preusker, Schenk, Otto, Hiesinger, O’Brien, Marchi,
  Sizemore, Hughson, Chilton, Bland, Byrne, Schorghofer, Platz, Jaumann,
  Roatsch, Sykes, Nathues, Sanctis, Raymond, \& Russell}]{Buczkowski2016}
Buczkowski, D.~L., Schmidt, B.~E., Williams, D.~A., {et~al.} 2016,
  \bibinfo{title}{The geomorphology of Ceres,} Science, 353,
  \dodoi{10.1126/science.aaf4332}

\bibitem[{D.~L. Buczkowski {et~al.}(2019)Buczkowski, Scully, Quick,
  Castillo-Rogez, Schenk, Park, Preusker, Jaumann, Raymond, \&
  Russell}]{Buczkowski2019}
Buczkowski, D.~L., Scully, J.~E., Quick, L., {et~al.} 2019,
  \bibinfo{title}{Tectonic analysis of fracturing associated with occator
  crater,} Icarus, 320, 49, \dodoi{10.1016/j.icarus.2018.05.012}

\bibitem[{J.~C. Castillo-Rogez {et~al.}(2019)Castillo-Rogez, Hesse, Formisano,
  Sizemore, Bland, Ermakov, \& Fu}]{Castillo-Rogez2019}
Castillo-Rogez, J.~C., Hesse, M.~A., Formisano, M., {et~al.} 2019,
  \bibinfo{title}{Conditions for the Long-Term Preservation of a Deep Brine
  Reservoir in Ceres,} Geophysical Research Letters, 46, 1963,
  \dodoi{10.1029/2018GL081473}

\bibitem[{H.~T. Chilton {et~al.}(2019)Chilton, Schmidt, Duarte, Ferrier,
  Hughson, Scully, Wray, Sizemore, Nathues, Platz, Schorghofer, Schenk, Landis,
  Bland, Byrne, Russell, \& Raymond}]{Chilton2019}
Chilton, H.~T., Schmidt, B.~E., Duarte, K., {et~al.} 2019,
  \bibinfo{title}{Landslides on Ceres: Inferences Into Ice Content and Layering
  in the Upper Crust,} Journal of Geophysical Research: Planets, 124, 1512,
  \dodoi{10.1029/2018JE005634}

\bibitem[{J.-P. Combe {et~al.}(2016)Combe, McCord, Tosi, Ammannito, Carrozzo,
  Sanctis, Raponi, Byrne, Landis, Hughson, Raymond, \& Russell}]{Combe2016}
Combe, J.-P., McCord, T.~B., Tosi, F., {et~al.} 2016, \bibinfo{title}{Detection
  of local H<sub>2</sub>O exposed at the surface of Ceres,} Science, 353,
  aaf3010, \dodoi{10.1126/science.aaf3010}

\bibitem[{E.~S. Costello {et~al.}(2020)Costello, Ghent, Hirabayashi, \&
  Lucey}]{Costello2020}
Costello, E.~S., Ghent, R.~R., Hirabayashi, M., \& Lucey, P.~G. 2020,
  \bibinfo{title}{Impact Gardening as a Constraint on the Age, Source, and
  Evolution of Ice on Mercury and the Moon,} Journal of Geophysical Research:
  Planets, 125, e2019JE006172, \dodoi{10.1029/2019JE006172}

\bibitem[{E.~S. Costello {et~al.}(2021)Costello, Ghent, \&
  Lucey}]{Costello2021}
Costello, E.~S., Ghent, R.~R., \& Lucey, P.~G. 2021, \bibinfo{title}{{Impact
  Gardening on Ceres},} Geophysical Research Letters, 48, e2021GL092960,
  \dodoi{10.1029/2021GL092960}

\bibitem[{ {Crater Analysis Techniques Working Group}(1979){Crater Analysis
  Techniques Working Group}}]{CraterAnalysis1979}
{Crater Analysis Techniques Working Group}. 1979, \bibinfo{title}{Standard
  techniques for presentation and analysis of crater size-frequency data,}
  Icarus, 37, \dodoi{10.1016/0019-1035(79)90009-5}

\bibitem[{K.~D. Duarte {et~al.}(2019)Duarte, Schmidt, Chilton, Hughson,
  Sizemore, Ferrier, Buffo, Scully, Nathues, Platz, Landis, Byrne, Bland,
  Russell, \& Raymond}]{Duarte2019}
Duarte, K.~D., Schmidt, B.~E., Chilton, H.~T., {et~al.} 2019,
  \bibinfo{title}{Landslides on Ceres: Diversity and Geologic Context,} Journal
  of Geophysical Research: Planets, 124, 3329, \dodoi{10.1029/2018JE005673}

\bibitem[{A.~I. Ermakov {et~al.}(2019)Ermakov, Kreslavsky, Scully, Hughson, \&
  Park}]{Ermakov2019}
Ermakov, A.~I., Kreslavsky, M.~A., Scully, J. E.~C., Hughson, K. H.~G., \&
  Park, R.~S. 2019, \bibinfo{title}{Surface Roughness and Gravitational Slope
  Distributions of Vesta and Ceres,} Journal of Geophysical Research: Planets,
  124, 14, \dodoi{10.1029/2018JE005813}

\bibitem[{C.~I. Fassett \& B.~J. Thomson(2014)Fassett \& Thomson}]{Fassett2014}
Fassett, C.~I., \& Thomson, B.~J. 2014, \bibinfo{title}{Crater degradation on
  the lunar maria: Topographic diffusion and the rate of erosion on the Moon,}
  Journal of Geophysical Research: Planets, 119, 2255,
  \dodoi{10.1002/2014JE004698}

\bibitem[{C.~I. Fassett {et~al.}(2022)Fassett, Beyer, Deutsch, Hirabayashi,
  Leight, Mahanti, Nypaver, Thomson, \& Minton}]{Fassett2022}
Fassett, C.~I., Beyer, R.~A., Deutsch, A.~N., {et~al.} 2022,
  \bibinfo{title}{Topographic Diffusion Revisited: Small Crater Lifetime on the
  Moon and Implications for Volatile Exploration,} Journal of Geophysical
  Research: Planets, 127, \dodoi{10.1029/2022JE007510}

\bibitem[{M. Formisano {et~al.}(2018)Formisano, Federico, De Sanctis, Frigeri,
  Magni, Raponi, \& Tosi}]{Formisano2018}
Formisano, M., Federico, C., De Sanctis, M.~C., {et~al.} 2018,
  \bibinfo{title}{Thermal Stability of Water Ice in Ceres' Craters: The Case of
  Juling Crater,} Journal of Geophysical Research: Planets, 123, 2445,
  \dodoi{10.1029/2017JE005417}

\bibitem[{A. Frigeri {et~al.}(2018)Frigeri, Schmedemann, Williams, Chemin,
  Mirino, Nass, Carrozzo, Castillo-Rogez, Buczkowski, Scully, Park, Crown,
  Mest, Federico, Ammannito, Sanctis, Raymond, \& Russell}]{Frigeri2018}
Frigeri, A., Schmedemann, N., Williams, D., {et~al.} 2018, \bibinfo{title}{The
  geology of the Nawish quadrangle of Ceres: The rim of an ancient basin,}
  Icarus, 316, \dodoi{10.1016/j.icarus.2018.08.015}

\bibitem[{R.~R. Fu {et~al.}(2017)Fu, Ermakov, Marchi, Castillo-Rogez, Raymond,
  Hager, Zuber, King, Bland, Sanctis, Preusker, Park, \& Russell}]{Fu2017}
Fu, R.~R., Ermakov, A.~I., Marchi, S., {et~al.} 2017, \bibinfo{title}{The
  interior structure of Ceres as revealed by surface topography,} Earth and
  Planetary Science Letters, 476, 153, \dodoi{10.1016/j.epsl.2017.07.053}

\bibitem[{S. Gou {et~al.}(2018)Gou, Yue, Di, \& Liu}]{Gou2018}
Gou, S., Yue, Z., Di, K., \& Liu, Z. 2018, \bibinfo{title}{A global catalogue
  of Ceres impact craters ≥ 1 km and preliminary analysis,} Icarus,
  302, 296, \dodoi{10.1016/j.icarus.2017.11.028}

\bibitem[{W.~K. Hartmann(1984)Hartmann}]{Hartmann1984}
Hartmann, W.~K. 1984, \bibinfo{title}{Does crater “saturation equilibrium”
  occur in the solar system?} Icarus, 60, 56,
  \dodoi{10.1016/0019-1035(84)90138-6}

\bibitem[{T. Heyer {et~al.}(2023)Heyer, Iqbal, Oetting, Hiesinger, van~der
  Bogert, \& Schmedemann}]{Heyer2023}
Heyer, T., Iqbal, W., Oetting, A., {et~al.} 2023, \bibinfo{title}{A comparative
  analysis of global lunar crater catalogs using OpenCraterTool – An open
  source tool to determine and compare crater size-frequency measurements,}
  Planetary and Space Science, 231, 105687, \dodoi{10.1016/j.pss.2023.105687}

\bibitem[{H. Hiesinger {et~al.}(2016)Hiesinger, Marchi, Schmedemann, Schenk,
  Pasckert, Neesemann, O’Brien, Kneissl, Ermakov, Fu, Bland, Nathues, Platz,
  Williams, Jaumann, Castillo-Rogez, Ruesch, Schmidt, Park, Preusker,
  Buczkowski, Russell, \& Raymond}]{Hiesinger2016}
Hiesinger, H., Marchi, S., Schmedemann, N., {et~al.} 2016,
  \bibinfo{title}{Cratering on Ceres: Implications for its crust and
  evolution,} Science, 353, \dodoi{10.1126/science.aaf4759}

\bibitem[{M. Hirabayashi {et~al.}(2024)Hirabayashi, Fassett, Costello, \&
  Minton}]{Hirabayashi2024}
Hirabayashi, M., Fassett, C.~I., Costello, E.~S., \& Minton, D.~A. 2024,
  \bibinfo{title}{Crater Equilibrium State Characterization given Crater
  Production from a Single Power Law,} The Planetary Science Journal, 5, 250,
  \dodoi{10.3847/psj/ad8883}

\bibitem[{M. Hirabayashi {et~al.}(2018)Hirabayashi, Howl, Fassett, Soderblom,
  Minton, \& Melosh}]{Hirabayashi2018}
Hirabayashi, M., Howl, B.~A., Fassett, C.~I., {et~al.} 2018,
  \bibinfo{title}{The Role of Breccia Lenses in Regolith Generation From the
  Formation of Small, Simple Craters: Application to the Apollo 15 Landing
  Site,} Journal of Geophysical Research: Planets, 123, 527,
  \dodoi{10.1002/2017JE005377}

\bibitem[{M. Hirabayashi {et~al.}(2017)Hirabayashi, Minton, \&
  Fassett}]{Hirabayashi2017}
Hirabayashi, M., Minton, D.~A., \& Fassett, C.~I. 2017, \bibinfo{title}{An
  analytical model of crater count equilibrium,} Icarus, 289, 134–143,
  \dodoi{10.1016/j.icarus.2016.12.032}

\bibitem[{K.~H.~G. Hughson {et~al.}(2019)Hughson, Russell, Schmidt, Chilton,
  Sizemore, Schenk, \& Raymond}]{Hughson2019}
Hughson, K. H.~G., Russell, C.~T., Schmidt, B.~E., {et~al.} 2019,
  \bibinfo{title}{Fluidized Appearing Ejecta on Ceres: Implications for the
  Mechanical Properties, Frictional Properties, and Composition of its Shallow
  Subsurface,} Journal of Geophysical Research: Planets, 124, 1819,
  \dodoi{10.1029/2018JE005666}

\bibitem[{B.~A. Ivanov {et~al.}(2002)Ivanov, Neukum, Bottke, \&
  Hartmann}]{Ivanov2002}
Ivanov, B.~A., Neukum, G., Bottke, W.~F., \& Hartmann, W.~K. 2002, The
  Comparison of Size-Frequency Distributions of Impact Craters and Asteroids
  and the Planetary Cratering Rate (The University of Arizona Press), 89--102,
  \dodoi{10.2307/j.ctv1v7zdn4.13}

\bibitem[{K. Krohn {et~al.}(2018)Krohn, Jaumann, Otto, Schulzeck, Neesemann,
  Nass, Stephan, Tosi, Wagner, Zambon, von~der Gathen, Williams, Buczkowski,
  Sanctis, Kersten, Matz, Mest, Pieters, Preusker, Roatsch, Scully, Russell, \&
  Raymond}]{Krohn2018}
Krohn, K., Jaumann, R., Otto, K., {et~al.} 2018, \bibinfo{title}{The unique
  geomorphology and structural geology of the Haulani crater of dwarf planet
  Ceres as revealed by geological mapping of equatorial quadrangle Ac-6
  Haulani,} Icarus, 316, 84, \dodoi{10.1016/j.icarus.2017.09.014}

\bibitem[{G. Kurt~Menke {et~al.}(2016)Kurt~Menke, Smith~Jr, Pirelli, John
  Van~Hoesen, {et~al.}}]{Kurt2016}
Kurt~Menke, G., Smith~Jr, R., Pirelli, L., John Van~Hoesen, G., {et~al.} 2016,
  Mastering QGIS (Packt Publishing Ltd)

\bibitem[{M.~E. Landis {et~al.}(2017)Landis, Byrne, Schörghofer, Schmidt,
  Hayne, Castillo-Rogez, Sykes, Combe, Ermakov, Prettyman, Raymond, \&
  Russell}]{Landis2017}
Landis, M.~E., Byrne, S., Schörghofer, N., {et~al.} 2017,
  \bibinfo{title}{Conditions for Sublimating Water Ice to Supply Ceres'
  Exosphere,} Journal of Geophysical Research: Planets, 122, 1984,
  \dodoi{10.1002/2017JE005335}

\bibitem[{S. Marchi {et~al.}(2016)Marchi, Ermakov, Raymond, Fu, O'Brien, Bland,
  Ammannito, Sanctis, Bowling, Schenk, Scully, Buczkowski, Williams, Hiesinger,
  \& Russell}]{Marchi2016}
Marchi, S., Ermakov, A.~I., Raymond, C.~A., {et~al.} 2016, \bibinfo{title}{The
  missing large impact craters on Ceres,} Nature Communications, 7,
  \dodoi{10.1038/ncomms12257}

\bibitem[{A.~S. McEwen \& E.~B. Bierhaus(2006)McEwen \& Bierhaus}]{McEwen2006}
McEwen, A.~S., \& Bierhaus, E.~B. 2006, \bibinfo{title}{THE IMPORTANCE OF
  SECONDARY CRATERING TO AGE CONSTRAINTS ON PLANETARY SURFACES,} Annual Review
  of Earth and Planetary Sciences, 34, 535,
  \dodoi{10.1146/annurev.earth.34.031405.125018}

\bibitem[{H.~J. Melosh(1989)Melosh}]{Melosh1989}
Melosh, H.~J. 1989, in Oxford Monographs on Geology and Geophysics, Vol.~11,
  Impact Cratering: A Geologic Process (New York: Oxford University Press)

\bibitem[{S. Mest {et~al.}(2016)Mest, Williams, Crown, Yingst, Buczkowski,
  Scully, Jaumann, Roatsch, Preusker, Nathues, Hoffmann, Schaefer, Raymond, \&
  Russell}]{Mest2016}
Mest, S., Williams, D.~A., Crown, D., {et~al.} 2016, in Geophysical Research
  Abstracts, Vol.~18, EGU General Assembly 2016

\bibitem[{D.~A. Minton {et~al.}(2019)Minton, Fassett, Hirabayashi, Howl, \&
  Richardson}]{Minton2019}
Minton, D.~A., Fassett, C.~I., Hirabayashi, M., Howl, B.~A., \& Richardson,
  J.~E. 2019, \bibinfo{title}{The equilibrium size-frequency distribution of
  small craters reveals the effects of distal ejecta on lunar landscape
  morphology,} Icarus, 326, 63, \dodoi{10.1016/j.icarus.2019.02.021}

\bibitem[{D.~A. Minton \& R. Malhotra(2010)Minton \& Malhotra}]{Minton2010}
Minton, D.~A., \& Malhotra, R. 2010, \bibinfo{title}{{Dynamical erosion of the
  asteroid belt and implications for large impacts in the inner Solar System},}
  Icarus, 207, 744, \dodoi{10.1016/j.icarus.2009.12.008}

\bibitem[{D.~A. Minton {et~al.}(2015)Minton, Richardson, \&
  Fassett}]{Minton2015}
Minton, D.~A., Richardson, J.~E., \& Fassett, C.~I. 2015,
  \bibinfo{title}{{Re-examining the main asteroid belt as the primary source of
  ancient lunar craters},} Icarus, 247, 172,
  \dodoi{10.1016/j.icarus.2014.10.018}

\bibitem[{P.~E. Montalvo {et~al.}(2022)Montalvo, Christopher, Hirabayashi, \&
  Fassett}]{Montalvo2022}
Montalvo, P.~E., Christopher, H., Hirabayashi, M., \& Fassett, C. 2022, in
  Lunar and Planetary Science Conference

\bibitem[{A. Morbidelli {et~al.}(2010)Morbidelli, Brasser, Gomes, Levison, \&
  Tsiganis}]{Morbidelli2010}
Morbidelli, A., Brasser, R., Gomes, R., Levison, H.~F., \& Tsiganis, K. 2010,
  \bibinfo{title}{EVIDENCE FROM THE ASTEROID BELT FOR A VIOLENT PAST EVOLUTION
  OF JUPITER's ORBIT,} The Astronomical Journal, 140, 1391,
  \dodoi{10.1088/0004-6256/140/5/1391}

\bibitem[{A. Nathues {et~al.}(2016)Nathues, Sierks, Gutierrez-Marques, Ripken,
  Hall, Buettner, Schaefer, \& Christensen}]{Nathues2016FC2Ceres}
Nathues, A., Sierks, H., Gutierrez-Marques, P., {et~al.} 2016,
  \bibinfo{title}{DAWN FC2 Calibrated Ceres Images V1.0,}, Dataset ID:
  DAWN-A-FC2-3-RDR-CERES-IMAGES-V1.0 NASA Planetary Data System

\bibitem[{A. Nathues {et~al.}(2017)Nathues, Platz, Thangjam, Hoffmann, Mengel,
  Cloutis, Corre, Reddy, Kallisch, \& Crown}]{Nathues2017}
Nathues, A., Platz, T., Thangjam, G., {et~al.} 2017, \bibinfo{title}{Evolution
  of Occator Crater on (1) Ceres,} The Astronomical Journal, 153, 112,
  \dodoi{10.3847/1538-3881/153/3/112}

\bibitem[{A. Nathues {et~al.}(2020)Nathues, Schmedemann, Thangjam, Pasckert,
  Mengel, Castillo-Rogez, Cloutis, Hiesinger, Hoffmann, Corre, Li, Pieters,
  Raymond, Reddy, Ruesch, \& Williams}]{Nathues2020}
Nathues, A., Schmedemann, N., Thangjam, G., {et~al.} 2020,
  \bibinfo{title}{Recent cryovolcanic activity at Occator crater on Ceres,}
  Nature Astronomy, 4, 794, \dodoi{10.1038/s41550-020-1146-8}

\bibitem[{D. Nesvorný {et~al.}(2017)Nesvorný, Roig, \& Bottke}]{Nesvorny2017}
Nesvorný, D., Roig, F., \& Bottke, W.~F. 2017, \bibinfo{title}{{Modeling the
  Historical Flux of Planetary Impactors},} The Astronomical Journal, 153, 103,
  \dodoi{10.3847/1538-3881/153/3/103}

\bibitem[{G. Neukum {et~al.}(2001)Neukum, Ivanov, \& Hartmann}]{Neukum2001}
Neukum, G., Ivanov, B., \& Hartmann, W. 2001, \bibinfo{title}{Cratering Records
  in the Inner Solar System in Relation to the Lunar Reference System,} Space
  Science Reviews, 96, 55–86, \dodoi{10.1023/a:1011989004263}

\bibitem[{D.~P. O'Brien {et~al.}(2014)O'Brien, Marchi, Morbidelli, Bottke,
  Schenk, Russell, \& Raymond}]{OBrien2014}
O'Brien, D.~P., Marchi, S., Morbidelli, A., {et~al.} 2014,
  \bibinfo{title}{Constraining the cratering chronology of Vesta,} Planetary
  and Space Science, 103, 131, \dodoi{10.1016/j.pss.2014.05.013}

\bibitem[{K.~A. Otto {et~al.}(2019)Otto, Marchi, Trowbridge, Melosh, \&
  Sizemore}]{Otto2019}
Otto, K.~A., Marchi, S., Trowbridge, A., Melosh, H.~J., \& Sizemore, H.~G.
  2019, \bibinfo{title}{Ceres Crater Degradation Inferred From Concentric
  Fracturing,} Journal of Geophysical Research: Planets, 124, 1188,
  \dodoi{10.1029/2018JE005660}

\bibitem[{J. Pasckert {et~al.}(2018)Pasckert, Hiesinger, Ruesch, Williams,
  Naß, Kneissl, Mest, Buczkowski, Scully, Schmedemann, Jaumann, Roatsch,
  Preusker, Nathues, Hoffmann, Schäfer, Sanctis, Raymond, \&
  Russell}]{Pasckert2018}
Pasckert, J., Hiesinger, H., Ruesch, O., {et~al.} 2018,
  \bibinfo{title}{Geologic mapping of the Ac-2 Coniraya quadrangle of Ceres
  from NASA's Dawn mission: Implications for a heterogeneously composed crust,}
  Icarus, 316, 28, \dodoi{10.1016/j.icarus.2017.06.015}

\bibitem[{T. Platz {et~al.}(2016)Platz, Nathues, Schorghofer, Preusker,
  Mazarico, Schr{\"o}der, Byrne, Kneissl, Schmedemann, Combe, Sch{\"a}fer,
  Thangjam, Hoffmann, Gutierrez-Marques, Landis, Dietrich, Ripken, Matz, \&
  Russell}]{Platz2016}
Platz, T., Nathues, A., Schorghofer, N., {et~al.} 2016, \bibinfo{title}{Surface
  water-ice deposits in the northern shadowed regions of Ceres,} Nature
  Astronomy, 1, 0007

\bibitem[{P. Pokorný {et~al.}(2021)Pokorný, Mazarico, \&
  Schorghofer}]{Pokorny2021}
Pokorný, P., Mazarico, E., \& Schorghofer, N. 2021, \bibinfo{title}{Erosion of
  volatiles by micro-meteoroid bombardment on Ceres, and comparison to the Moon
  and Mercury,}

\bibitem[{R. Povilaitis {et~al.}(2018)Povilaitis, Robinson, van~der Bogert,
  Hiesinger, Meyer, \& Ostrach}]{Povilaitis2018}
Povilaitis, R., Robinson, M., van~der Bogert, C., {et~al.} 2018,
  \bibinfo{title}{Crater density differences: Exploring regional resurfacing,
  secondary crater populations, and crater saturation equilibrium on the moon,}
  Planetary and Space Science, 162, 41, \dodoi{10.1016/j.pss.2017.05.006}

\bibitem[{ {QGIS Development Team}(2021){QGIS Development
  Team}}]{QGIS_software}
{QGIS Development Team}. 2021, QGIS Geographic Information System, QGIS
  Association.
\newblock \url{https://www.qgis.org}

\bibitem[{L.~C. Quick {et~al.}(2019)Quick, Buczkowski, Ruesch, Scully,
  Castillo-Rogez, Raymond, Schenk, Sizemore, \& Sykes}]{Quick2019}
Quick, L.~C., Buczkowski, D.~L., Ruesch, O., {et~al.} 2019, \bibinfo{title}{A
  Possible Brine Reservoir Beneath Occator Crater: Thermal and Compositional
  Evolution and Formation of the Cerealia Dome and Vinalia Faculae,} Icarus,
  320, 119, \dodoi{10.1016/j.icarus.2018.07.016}

\bibitem[{J.~E. Richardson(2009)Richardson}]{Richardson2009}
Richardson, J.~E. 2009, \bibinfo{title}{Cratering saturation and equilibrium: A
  new model looks at an old problem,} Icarus, 204, 697,
  \dodoi{10.1016/j.icarus.2009.07.029}

\bibitem[{C. Riedel {et~al.}(2020)Riedel, Minton, Michael, Orgel, van~der
  Bogert, \& Hiesinger}]{Riedel2020}
Riedel, C., Minton, D.~A., Michael, G., {et~al.} 2020,
  \bibinfo{title}{Degradation of Small Simple and Large Complex Lunar Craters:
  Not a Simple Scale Dependence,} Journal of Geophysical Research: Planets,
  125, \dodoi{10.1029/2019JE006273}

\bibitem[{S.~J. Robbins {et~al.}(2025)Robbins, Kirchoff, \&
  Ostrach}]{Robbins2025}
Robbins, S.~J., Kirchoff, M.~R., \& Ostrach, L.~R. 2025, \bibinfo{title}{Crater
  Detection Dependence on Resolution, Incidence Angle, Emission Angle, and
  Phase Angle,} Geophysical Research Letters, 52, \dodoi{10.1029/2024GL110570}

\bibitem[{S.~J. Robbins {et~al.}(2014)Robbins, Antonenko, Kirchoff, Chapman,
  Fassett, Herrick, Singer, Zanetti, Lehan, Huang, \& Gay}]{Robbins2014}
Robbins, S.~J., Antonenko, I., Kirchoff, M.~R., {et~al.} 2014,
  \bibinfo{title}{The variability of crater identification among expert and
  community crater analysts,} Icarus, 234, 109,
  \dodoi{10.1016/j.icarus.2014.02.022}

\bibitem[{O. Ruesch {et~al.}(2016)Ruesch, Platz, Schenk, McFadden,
  Castillo-Rogez, Quick, Byrne, Preusker, O’Brien, Schmedemann, Williams, Li,
  Bland, Hiesinger, Kneissl, Neesemann, Schaefer, Pasckert, Schmidt,
  Buczkowski, Sykes, Nathues, Roatsch, Hoffmann, Raymond, \&
  Russell}]{Ruesch2016}
Ruesch, O., Platz, T., Schenk, P., {et~al.} 2016, \bibinfo{title}{Cryovolcanism
  on Ceres,} Science, 353, aaf4286, \dodoi{10.1126/science.aaf4286}

\bibitem[{C.~T. Russell \& C.~A. Raymond(2011)Russell \& Raymond}]{Russell2011}
Russell, C.~T., \& Raymond, C.~A. 2011, \bibinfo{title}{The Dawn Mission to
  Vesta and Ceres,} Space Science Reviews, 163, 3,
  \dodoi{10.1007/s11214-011-9836-2}

\bibitem[{C.~T. Russell {et~al.}(2016)Russell, Raymond, Ammannito, Buczkowski,
  Sanctis, Hiesinger, Jaumann, Konopliv, McSween, Nathues, Park, Pieters,
  Prettyman, McCord, McFadden, Mottola, Zuber, Joy, Polanskey, Rayman,
  Castillo-Rogez, Chi, Combe, Ermakov, Fu, Hoffmann, Jia, King, Lawrence, Li,
  Marchi, Preusker, Roatsch, Ruesch, Schenk, Villarreal, \&
  Yamashita}]{Russell2016}
Russell, C.~T., Raymond, C.~A., Ammannito, E., {et~al.} 2016,
  \bibinfo{title}{Dawn arrives at ceres: Exploration of a small, volatile-rich
  world,} Science, 353, \dodoi{10.1126/science.aaf4219}

\bibitem[{P. Schenk {et~al.}(2004)Schenk, Chapman, Zahnle, \&
  Moore}]{Schenk2004}
Schenk, P., Chapman, C.~R., Zahnle, K., \& Moore, J.~M. 2004, Ages and
  interiors: the cratering record of the Galilean satellites

\bibitem[{P. Schenk {et~al.}(2020)Schenk, Scully, Buczkowski, Sizemore,
  Schmidt, Pieters, Neesemann, O'Brien, Marchi, Williams, Nathues, De~Sanctis,
  Tosi, Russell, Castillo-Rogez, \& Raymond}]{Schenk2020}
Schenk, P., Scully, J., Buczkowski, D., {et~al.} 2020, \bibinfo{title}{Impact
  heat driven volatile redistribution at Occator crater on Ceres as a
  comparative planetary process,} Nature Communications, 11, 3679

\bibitem[{N. Schmedemann {et~al.}(2014)Schmedemann, Kneissl, Ivanov, Michael,
  Wagner, Neukum, Ruesch, Hiesinger, Krohn, Roatsch, Preusker, Sierks, Jaumann,
  Reddy, Nathues, Walter, Neesemann, Raymond, \& Russell}]{Schmedemann2014}
Schmedemann, N., Kneissl, T., Ivanov, B., {et~al.} 2014, \bibinfo{title}{The
  cratering record, chronology and surface ages of (4) Vesta in comparison to
  smaller asteroids and the ages of HED meteorites,} Planetary and Space
  Science, 103, 104, \dodoi{10.1016/j.pss.2014.04.004}

\bibitem[{B.~E. Schmidt {et~al.}(2017)Schmidt, Hughson, Chilton, Scully, Platz,
  Nathues, Sizemore, Bland, Byrne, Marchi, O'Brien, Schorghofer, Hiesinger,
  Jaumann, Pasckert, Lawrence, Buzckowski, Castillo-Rogez, Sykes, Schenk,
  DeSanctis, Mitri, Formisano, Li, Reddy, LeCorre, Russell, \&
  Raymond}]{Schmidt2017}
Schmidt, B.~E., Hughson, K. H.~G., Chilton, H.~T., {et~al.} 2017,
  \bibinfo{title}{Geomorphological evidence for ground ice on dwarf planet
  Ceres,} Nature Geoscience, 10, 338, \dodoi{10.1038/ngeo2936}

\bibitem[{B.~E. Schmidt {et~al.}(2020)Schmidt, Sizemore, Hughson, Duarte,
  Romero, Scully, Schenk, Buczkowski, Williams, Nathues, Udell, Castillo-Rogez,
  Raymond, \& Russell}]{Schmidt2020}
Schmidt, B.~E., Sizemore, H.~G., Hughson, K. H.~G., {et~al.} 2020,
  \bibinfo{title}{Post-impact cryo-hydrologic formation of small mounds and
  hills in Ceres's Occator crater,} Nature Geoscience, 13, 605,
  \dodoi{10.1038/s41561-020-0581-6}

\bibitem[{S.~E. Schr{\"o}der {et~al.}(2021)Schr{\"o}der, Poch, Ferrari,
  Angelis, Sultana, Potin, Beck, De~Sanctis, \& Schmitt}]{Schroder2021}
Schr{\"o}der, S.~E., Poch, O., Ferrari, M., {et~al.} 2021,
  \bibinfo{title}{Dwarf planet (1) Ceres surface bluing due to high porosity
  resulting from sublimation,} Nature Communications, 12, 274,
  \dodoi{10.1038/s41467-020-20494-5}

\bibitem[{J.~E. Scully {et~al.}(2017)Scully, Buczkowski, Neesemann, Williams,
  Mest, Raymond, Nass, Hughson, Kneissl, Pasckert, Ruesch, Frigeri, Marchi,
  Combe, Schmedemann, Schmidt, Chilton, Russell, Jaumann, Preusker, Roatsch,
  Hoffmann, Nathues, Schaefer, \& Ermakov}]{Scully2017}
Scully, J.~E., Buczkowski, D., Neesemann, A., {et~al.} 2017,
  \bibinfo{title}{Ceres’ Ezinu quadrangle: a heavily cratered region with
  evidence for localized subsurface water ice and the context of Occator
  crater,} Icarus, 316, 46, \dodoi{10.1016/j.icarus.2017.10.038}

\bibitem[{J.~E. Scully {et~al.}(2018)Scully, Buczkowski, Neesemann, Williams,
  Mest, Raymond, Nass, Hughson, Kneissl, Pasckert, Ruesch, Frigeri, Marchi,
  Combe, Schmedemann, Schmidt, Chilton, Russell, Jaumann, Preusker, Roatsch,
  Hoffmann, Nathues, Schaefer, \& Ermakov}]{Scully2018}
Scully, J.~E., Buczkowski, D., Neesemann, A., {et~al.} 2018,
  \bibinfo{title}{{Ceres’ Ezinu quadrangle: a heavily cratered region with
  evidence for localized subsurface water ice and the context of Occator
  crater},} Icarus, 316, 46, \dodoi{10.1016/j.icarus.2017.10.038}

\bibitem[{J.~E. Scully {et~al.}(2019{\natexlab{a}})Scully, Buczkowski, Raymond,
  Bowling, Williams, Neesemann, Schenk, Castillo-Rogez, \&
  Russell}]{Scully2019a}
Scully, J.~E., Buczkowski, D.~L., Raymond, C.~A., {et~al.} 2019{\natexlab{a}},
  \bibinfo{title}{Ceres’ Occator crater and its faculae explored through
  geologic mapping,} Icarus, 320, 7, \dodoi{10.1016/j.icarus.2018.04.014}

\bibitem[{J.~E. Scully {et~al.}(2019{\natexlab{b}})Scully, Bowling, Bu,
  Buczkowski, Longobardo, Nathues, Neesemann, Palomba, Quick, Raponi, Ruesch,
  Schenk, Stein, Thomas, Russell, Castillo-Rogez, Raymond, \&
  Jaumann}]{Scully2019b}
Scully, J.~E., Bowling, T., Bu, C., {et~al.} 2019{\natexlab{b}},
  \bibinfo{title}{Synthesis of the special issue: The formation and evolution
  of Ceres’ Occator crater,} Icarus, 320, 213,
  \dodoi{10.1016/j.icarus.2018.08.029}

\bibitem[{H. Sierks {et~al.}(2011)Sierks, Keller, Jaumann, Michalik, Behnke,
  Bubenhagen, Büttner, Carsenty, Christensen, Enge, Fiethe, Marqués, Hartwig,
  Krüger, Kühne, Maue, Mottola, Nathues, Reiche, Richards, Roatsch,
  Schröder, Szemerey, \& Tschentscher}]{Sierks2011}
Sierks, H., Keller, H.~U., Jaumann, R., {et~al.} 2011, \bibinfo{title}{The Dawn
  Framing Camera,} Space Science Reviews, 163, 263,
  \dodoi{10.1007/s11214-011-9745-4}

\bibitem[{S.~K. Singh {et~al.}(2021)Singh, Bergantini, Zhu, Ferrari,
  De~Sanctis, De~Angelis, \& Kaiser}]{Singh2021}
Singh, S.~K., Bergantini, A., Zhu, C., {et~al.} 2021, \bibinfo{title}{Origin of
  ammoniated phyllosilicates on dwarf planet Ceres and asteroids,} Nature
  Communications, 12, 2690

\bibitem[{L.~A. Soderblom(1970)Soderblom}]{Soderblom1970}
Soderblom, L.~A. 1970, \bibinfo{title}{{A model for small-impact erosion
  applied to the lunar surface},} Journal of Geophysical Research (1896-1977),
  75, 2655, \dodoi{10.1029/JB075i014p02655}

\bibitem[{M.~M. Sori {et~al.}(2017)Sori, Byrne, Bland, Bramson, Ermakov,
  Hamilton, Otto, Ruesch, \& Russell}]{Sori2017}
Sori, M.~M., Byrne, S., Bland, M.~T., {et~al.} 2017, \bibinfo{title}{The
  vanishing cryovolcanoes of Ceres,} Geophysical Research Letters, 44, 1243,
  \dodoi{10.1002/2016GL072319}

\bibitem[{R. Strom {et~al.}(2018)Strom, Marchi, \& Malhotra}]{Strom2018}
Strom, R., Marchi, S., \& Malhotra, R. 2018, \bibinfo{title}{Ceres and the
  terrestrial planets impact cratering record,} Icarus, 302, 104,
  \dodoi{10.1016/j.icarus.2017.11.013}

\bibitem[{C.~L. Talkington {et~al.}(2022)Talkington, Hirabayashi, Montalvo,
  Deutsch, Fassett, Siegler, Shepherd, \& King}]{Talkington2022}
Talkington, C.~L., Hirabayashi, M., Montalvo, P.~E., {et~al.} 2022,
  \bibinfo{title}{Survival of Ancient Lunar Water Affected by Topographic
  Degradation of Old, Large Complex Craters,} Geophysical Research Letters, 49,
  \dodoi{10.1029/2022GL099241}

\bibitem[{K. Toyokawa {et~al.}(2022)Toyokawa, Haruyama, Hirata, Tanaka, \&
  Iwata}]{Toyokawa2022}
Toyokawa, K., Haruyama, J., Hirata, N., Tanaka, S., \& Iwata, T. 2022,
  \bibinfo{title}{Kilometer-scale crater size-frequency distributions on
  Ceres,} Icarus, 377, 114909, \dodoi{10.1016/j.icarus.2022.114909}

\bibitem[{Z. Xiao \& S.~C. Werner(2015)Xiao \& Werner}]{Xiao2015}
Xiao, Z., \& Werner, S.~C. 2015, \bibinfo{title}{Size‐frequency distribution
  of crater populations in equilibrium on the Moon,} Journal of Geophysical
  Research: Planets, 120, 2277, \dodoi{10.1002/2015JE004860}

\bibitem[{M. Zeilnhofer(2020)Zeilnhofer}]{Michael2020}
Zeilnhofer, M. 2020, PhD thesis, Northern Arizona University

\bibitem[{M.~F. Zeilnhofer \& N.~G. Barlow(2020)Zeilnhofer \&
  Barlow}]{Zeilnhofer2020}
Zeilnhofer, M.~F., \& Barlow, N.~G. 2020, in Lunar and Planetary Science
  Conference

\end{thebibliography}



\end{document}